 \documentclass[journal=jpccfh]{achemso}
 \usepackage[version=3]{mhchem} 
\usepackage{color}
\usepackage{verbatim}
\usepackage{mciteplus}
\usepackage{amsbsy}
\usepackage{amstext}
\usepackage{amssymb}
\usepackage{natmove}
\usepackage{caption}
\usepackage{float}
\usepackage{geometry}
\usepackage{setspace}
\usepackage{xkeyval}
\usepackage{graphicx}
\usepackage{subfig}
\usepackage{placeins}
\usepackage{longtable}
\usepackage{notes2bib}
\usepackage{natbib}
\usepackage{colortbl}
 
\DeclareGraphicsExtensions{.jpg, .pdf, .png}

\definecolor{JPCCBlue}{RGB}{34,80,169}
\definecolor{NLRed}{RGB}{215,24,30}
\definecolor{abstractcolor}{RGB}{255,243,201}
\definecolor{tablegray}{gray}{0.85}

\newcolumntype{M}{>{\centering\arraybackslash}m{1.6cm}}

 \usepackage[version=3]{mhchem} % Formula subscripts using \ce{}
 \usepackage[unicode=true,
  bookmarks=true,bookmarksnumbered=false,bookmarksopen=false,
  breaklinks=false,pdfborder={0 0 0},backref=false,colorlinks=true]
  {hyperref}
 \hypersetup{pdftitle={Large-scale first principles configuration interaction calculations of optical absorption in aluminum clusters},
  pdfauthor={Ravindra Shinde, Alok Shukla},
  pdfsubject={Computational Condensed Matter},
  unicode=false,pdftoolbar=true,pdfmenubar=true,pdffitwindow=true,pdfstartview={FitH},pdfnewwindow=true,pdfcreator={RevTeX},linkcolor=red,citecolor=blue,urlcolor=blue}
 
 \makeatletter

\providecommand{\tabularnewline}{\\}

\usepackage{titlesec}
\titleformat{\section}{\bfseries\sffamily\color{JPCCBlue}}{\thesection.~}{0pt}{}
\titleformat{\subsection}[runin]{\bfseries\sffamily\normalsize}{\indent\thesubsection.~}{0pt}{}[.]
\titlespacing{\subsection}{0pt}{0pt}{*1}
\newcommand\nofootnote[1]{%
  \begingroup
  \renewcommand\thefootnote{}\footnote{#1}%
  \addtocounter{footnote}{-1}%
  \endgroup
}

   \@ifundefined{showcaptionsetup}{}{%
    \PassOptionsToPackage{caption=false}{subfig}}
   \usepackage{subfig}

 \makeatother

\title{Large-scale first principles configuration interaction calculations of optical absorption in aluminum clusters}

\author{Ravindra Shinde$^{a}$}
\affiliation{Department of Physics, Indian Institute of Technology Bombay, Mumbai, Maharashtra 400076, INDIA.}
\email{ravindra.shinde@iitb.ac.in}
  \phone{+91 (0)22 25764558}
  \fax{+91 (0)22 25767552}

\author{Alok Shukla$^{b}$}
\email{shukla@phy.iitb.ac.in}
\affiliation{Department of Physics, Indian Institute of Technology Bombay, Mumbai,
Maharashtra 400076, INDIA.}

\begin{document}

\nofootnote{\dag~Electronic Supplementary Information (ESI) available: 
A detailed information about wave functions of excited states contributing to various photoabsorption peaks is presented 
in the supplementary information (Table I through IX). See DOI: 10.1039/b000000x/}
%Please use \dag to cite the ESI in the main text of the article.
%If you article does not have ESI please remove the the \dag symbol from the title and the above footnotetext.
\nofootnote{\textit{$^{a}$Department of Physics, Indian Institute of Technology Bombay, Mumbai, Maharashtra 400076, India. Fax: +91 22 2576 7552; Tel: +91 22 2576 4558; E-mail: ravindra.shinde@iitb.ac.in}}
\nofootnote{\textit{$^{b}$Department of Physics, Indian Institute of Technology Bombay, Mumbai, Maharashtra 400076, India. Fax: +91 22 2576 7552; Tel: +91 22 2576 7576; E-mail: shukla@phy.iitb.ac.in}}

\begin{abstract}
We report the linear optical absorption spectra of aluminum clusters
Al$_{n}$ (n=2--5) involving valence transitions, computed using the  large-scale all-electron  
 configuration interaction (CI) methodology. Several 
low-lying isomers of each cluster were considered,
and their geometries were optimized at the coupled-cluster singles
doubles (CCSD) level of theory. With these optimized ground-state
geometries, excited states of different clusters were computed using
the multi-reference singles-doubles configuration-interaction (MRSDCI)
approach, which includes electron correlation effects at a sophisticated
level. These CI wave functions were used to compute the transition
dipole matrix elements connecting the ground and various excited states
of different clusters, and thus their photoabsorption
spectra. The convergence of our results with respect to the basis
sets, and the size of the CI expansion, was carefully examined.
Our results were found to be significantly different as compared to those obtained using time-dependent
density functional theory (TDDFT) [Deshpande \textit{et al.  Phys. Rev. B}, 2003, \textbf{68}, 035428].
When compared to available experimental data for the isomers of  Al$_{2}$ and Al$_{3}$, 
our results are in very good agreement as far as important peak positions are concerned. 
 The contribution of configurations to many body wavefunction of various
excited states suggests that in most cases optical excitations involved are collective, 
and plasmonic in nature.
\end{abstract}

\section{Introduction}

Metal clusters are promising candidates in the era of nanotechnology.
The reason behind growing interest in clusters lies in their interesting
properties and a vast variety of potential technological applications \cite{alonso_book,clustnano_book,julius_book,deheer_rmp,na_dehaar_prl}.
Moreover, simple theoretical models can be exploited to describe their properties.

Various jellium models have successfully described electronic structures
of alkali metal clusters, because alkali metals have free valence
electrons \cite{deheer_rmp}. This beautifully explains the higher abundance of certain
clusters. However, in case of aluminum clusters, the experimental
results often provide conflicting evidence about the size at which the jellium model would work \cite{nonjellium-to-jellium, wang_sp_hybrid_prl}. 
The theoretical explanation also depends on the valency of aluminum
atoms considered. Since s--p orbital energy separation in aluminum atom is 4.99
eV, and it decreases with the cluster size, the valency should be
changed from one to three \cite{rao_jena_ele_struct_al}. Perturbed jellium model, which takes orbital anisotropy
into account, has successfully explained the mass abundance of aluminum clusters \cite{upton_elec_struct_small_al,free_ele_metal_cluster_clemenger}.

Shell structure and s--p hybridization in anionic aluminum clusters
were probed using photoelectron spectroscopy by Gantef\"{o}r and Eberhardt \cite{Gante_shell_struct_al},
and Li \emph{et al}. \cite{wang_sp_hybrid_prl}. Evolution of electronic
structure and other properties of aluminum clusters has been studied
in many reports \cite{rao_jena_ele_struct_al,reinhart_big_al_cluster,sachdev_dft_al_small,
truhler_theory_validation_jpcb,upton_elec_struct_small_al,wang_sp_hybrid_prl,whetten_struct_bind_prb,
cox-al3-experiment,tse-al2-al3-1988,tse-al3-jcp-1990,jones_simul_anneal_al,jones_struct_bind_prl,
manninen_ionization_al,martinez_all_vs_core,dft_al_dimer_jmolstruct,al3_al3_anion_excited_II_jcp,
drebov_and_ahlrichs,mojtaba_stat_dyna_pol_al}.
Structural properties of aluminum clusters were studied using density
functional theory by Rao and Jena \cite{rao_jena_ele_struct_al}.
An all electron and model core potential study of various Al clusters was carried out by
Martinez \emph{et al}. \cite{martinez_all_vs_core}. Upton performed
chemisorption calculations on aluminum clusters and reported that
Al$_{6}$ is the smallest cluster that will absorb H$_{2}$.\cite{upton_elec_struct_small_al}
DFT alongwith molecular dynamics were used to study electronic and
structural properties of aluminum clusters \cite{jones_simul_anneal_al}. Among more recent works, Drebov and Ahlrichs~\cite{drebov_and_ahlrichs} presented a very detailed and systematic study of geometrical structure and electronic properties of large Al clusters ranging from  Al$_{23}$ to Al$_{34}$, and their anions and cations. Alipour and Mohajeri\cite{mojtaba_stat_dyna_pol_al} performed a comprehensive study of the electronic structure, ionization potential, and static and dynamic polarizabilities (at a fixed frequency) of clusters 
ranging from Al$_3$ to Al$_{31}$.

Although the photoabsorption in alkali metal clusters has been studied by many authors at various levels of 
theory \cite{deheer_rmp,yanno-frag-absorption-prl89, *na_opt_kappes_jcp, *optical_na_deheer,
*yanno-optical-response-pra91, *yanno-evolution-optical-prb93, *li_na_tddft, *yanno-surface-plasmon-prb98, *na_opt_bethe_salp}, however, very few theoretical calculations of the photoabsorption spectra in aluminum clusters exist.\cite{kanhere_prb_optical_al,smith_tunable_icosa}. 
As far experimental studies of optical absorption in aluminum clusters are concerned, several studies have been performed on  Al$_2$~\cite{ginter_al2_apj_63,absorption_spectra_al_dimer_jpc,spectra_jet_cooled_trimer_jcp,bondybey_expt_3piu_dimer} and Al$_3$.\cite{al3_al3_anion_excited_I_jcp, spectra_jet_cooled_trimer_jcp, spectra_trimer_ijms, infrared_spectra_boron_aluminum_cpl} Nevertheless, to the best of our knowledge, no experimental measurements of 
optical properties of larger aluminum clusters have been performed.
  
Conventional mass spectrometry only distinguishes clusters according
to the masses. Hence,  theoretical results can be coupled with the experimental
measurements of optical absorption, to distinguish between different
isomers of a cluster. This is important for clusters of increasing larger sizes for which several possible 
isomers exist. We have recently reported results of such calculation
on small boron clusters \cite{smallboron}. In this paper, we present results of systematic calculations of 
linear optical absorption involving transitions among valence states in various low-lying isomers of 
small aluminum clusters using \emph{ab initio} large-scale multi-reference singles doubles configuration 
interaction (MRSDCI) method. In our group, in the past we have successfully employed the MRSDCI approach to
 compute the photoabsorption spectra of a number of conjugated 
polymers,\cite{mrsd_jcp_09,mrsd_prb_02,mrsd_prb_05,mrsd_prb_07} and boron clusters.\cite{smallboron, borozene-sahu}
 Therefore, it is our 
intention in this work to test this approach on clusters made up of larger atoms, namely aluminum, and critically analyze its performance.  Furthermore, the nature of optical excitations involved in absorption has also been investigated by analyzing the wave functions of the excited states.

Upon comparing calculated optical absorption spectra of  Al$_2$ and Al$_3$, we find very good agreement 
with the available experimental data on important peaks. This suggests that the MRSDCI approach is equally effective for Al clusters, as it was, say, for boron clusters.\cite{smallboron, borozene-sahu} For larger clusters, for which no experimental data is available, we compare our results with the time-dependent 
density functional theory (TDDFT)   based calculations of Deshpande \emph{et al.}~\cite{kanhere_prb_optical_al} 
 corresponding to the minimum energy configurations, and find significant differences.

Remainder of the paper is organized as follows. Next section discusses
theoretical and computational details of the calculations, followed
by section \ref{sec:RESULTS-AND-DISCUSSION}, in which results are
presented and discussed. Conclusions and future directions are presented
in section \ref{sec:CONCLUSIONS-AND-OUTLOOK}. A detailed information about the nature of optical excitation, molecular orbitals of 
clusters, wave functions of excited states contributing to various photoabsorption peaks is presented in 
the supplementary information\cite{supplementary-material}.

\section{\label{sec:theory}Theoretical and Computational Details}

The geometry of various isomers were optimized using the size-consistent
coupled-cluster singles-doubles (CCSD) method, as implemented in the
\textsc{gaussian 09} package \cite{gaussian09}. A basis set of 6-311++G(2d,2p)
was used which was included in the \textsc{gaussian 09} package itself. This
basis set is optimized for the ground state calculations. 

\begin{figure*}[!t]
\centering
\subfloat[\textbf{Al$_{\mathbf{2}}$, D$_{\mathbf{\boldsymbol{\infty}h}}$,
$^{\mathbf{3}}\mathbf{\boldsymbol{\Pi_{u}}}$ \label{subfig:subfig-al2}}]{\includegraphics[width=2.8cm]{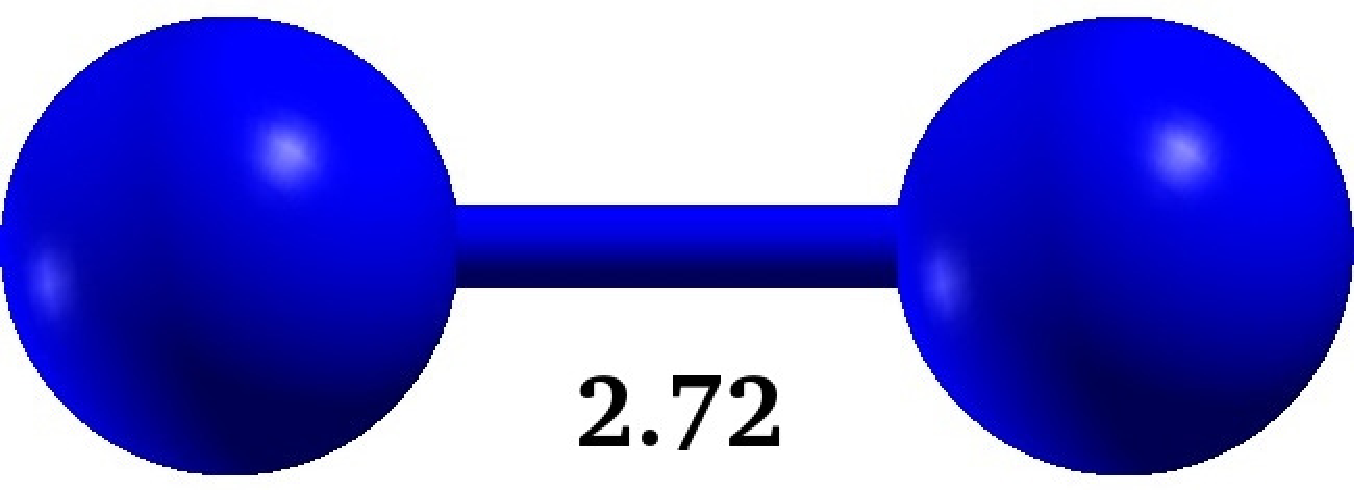}}\hfill 
\subfloat[\textbf{Al$_{\mathbf{2}}$, D$_{\mathbf{\boldsymbol{\infty}h}}$,
$^{\mathbf{3}}\mathbf{\boldsymbol{\Sigma_{g}}}$ \label{subfig:subfig-al2-meta}}]{\includegraphics[width=0.12\paperwidth]{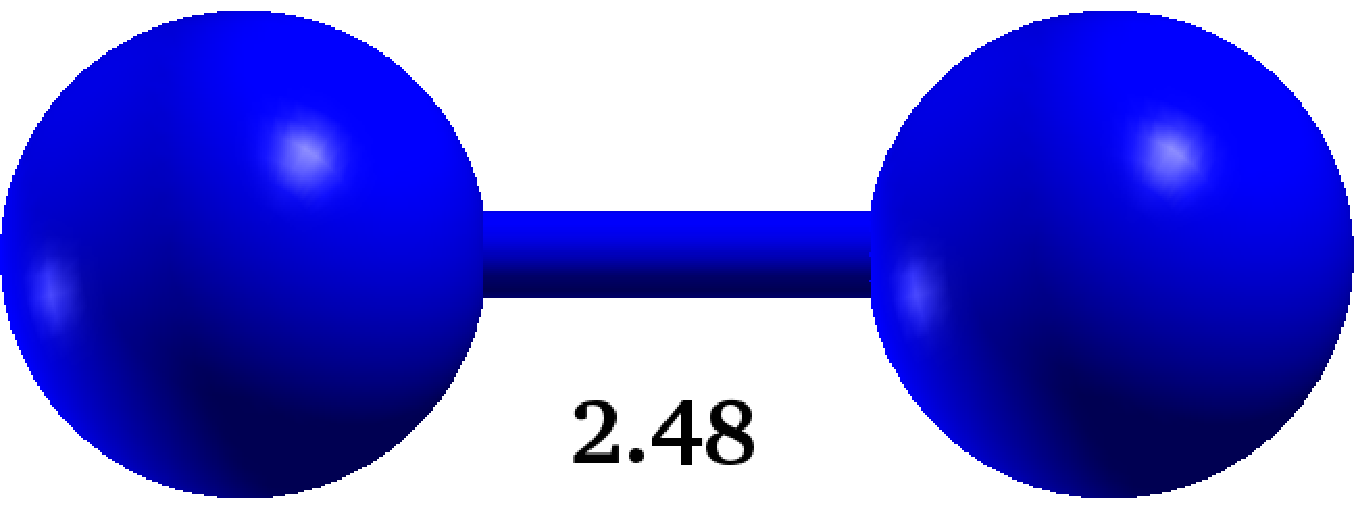}}
\hfill \subfloat[\textbf{Al$_{\boldsymbol{3}}$, D$_{\boldsymbol{3h}}$, $\boldsymbol{^{2}A_{1}^{'}}$}]{\includegraphics[width=0.12\paperwidth]{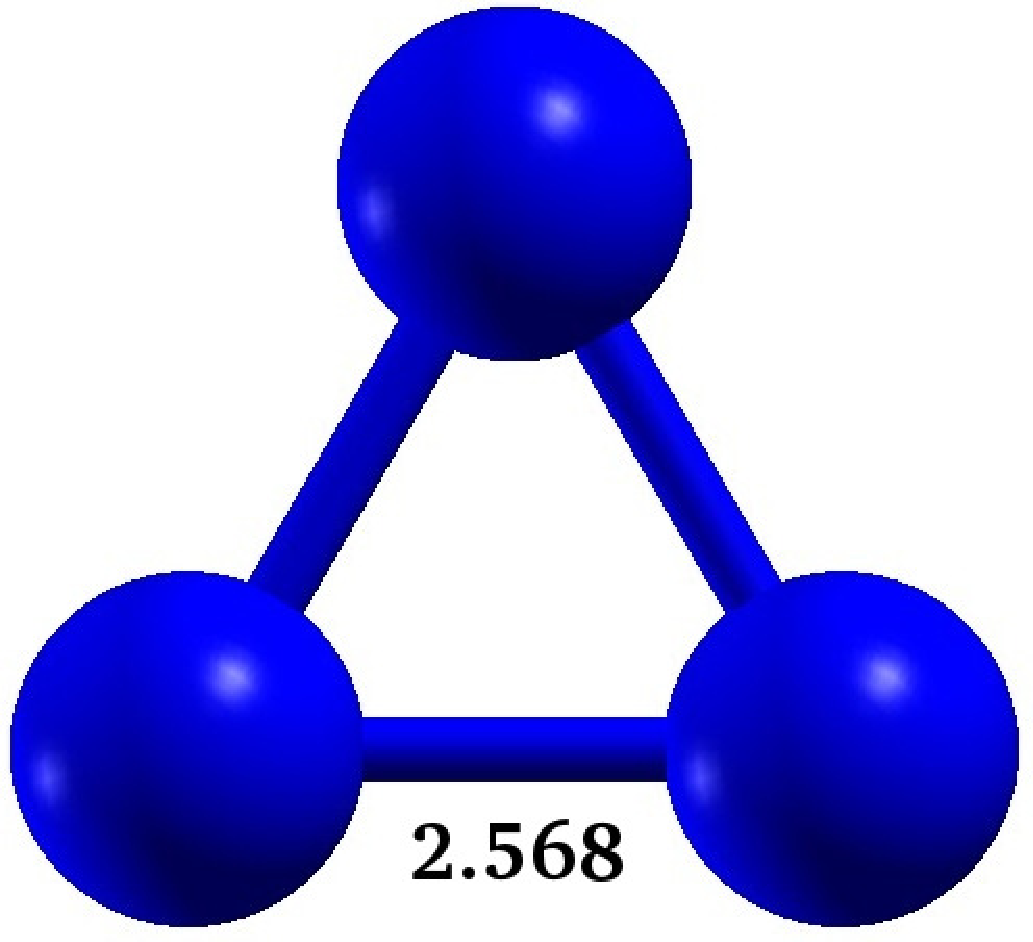}
}
\hfill \subfloat[\textbf{Al$_{\boldsymbol{3}}$, C}$_{\boldsymbol{2v}}$, $\boldsymbol{^{4}A_{2}}$]{\includegraphics[width=0.14\paperwidth]{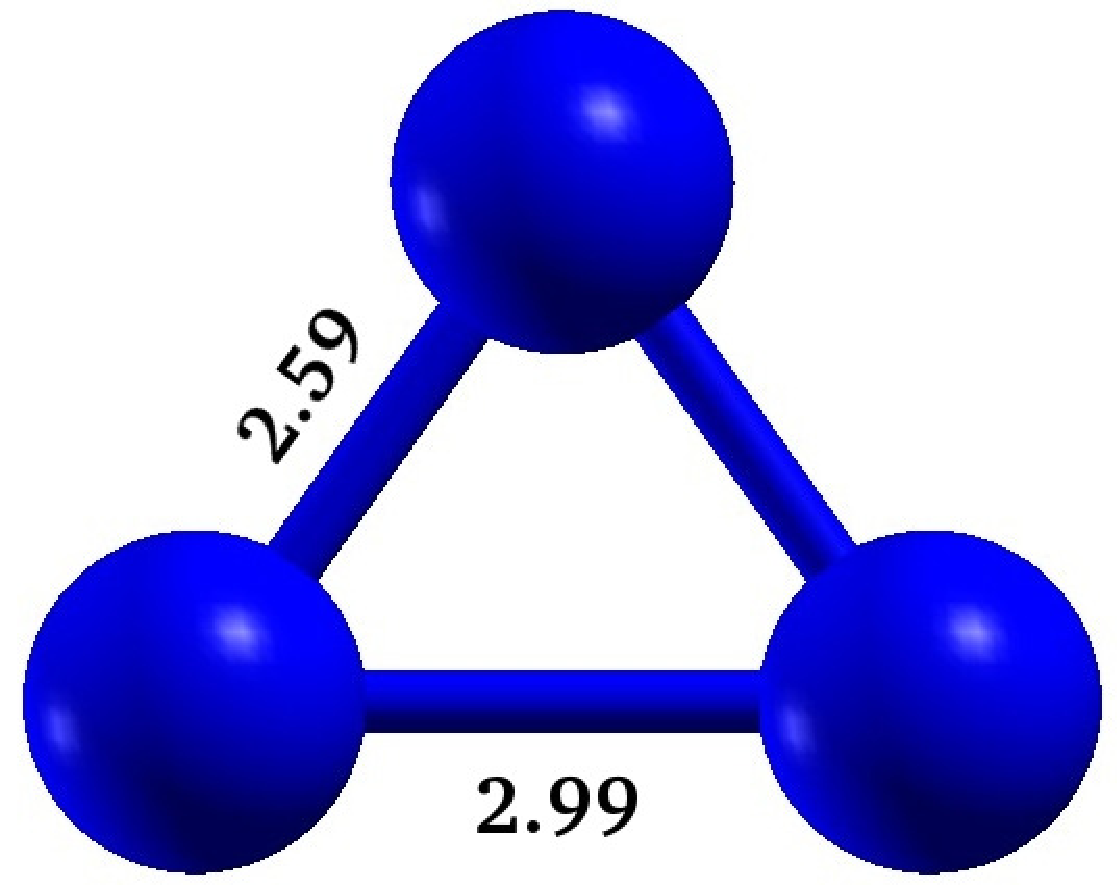}

}\hfill \subfloat[\textbf{Al$_{\boldsymbol{3}}$, D$_{\boldsymbol{\infty h}}$, $\boldsymbol{^{4}\Sigma_{u}}$}]{\includegraphics[width=0.2\paperwidth]{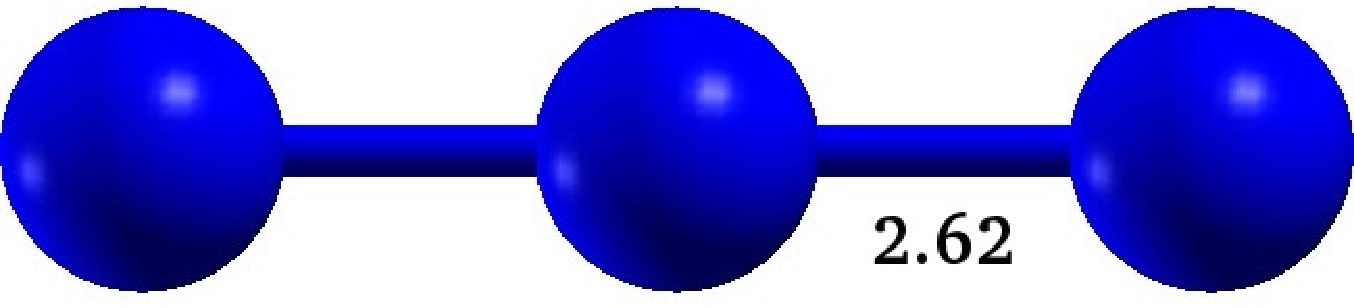}

}

\subfloat[\textbf{Al$_{\boldsymbol{4}}$, D$_{\boldsymbol{2h}}$, $\boldsymbol{^{3}B{}_{2g}}$}]{\includegraphics[width=0.17\paperwidth]{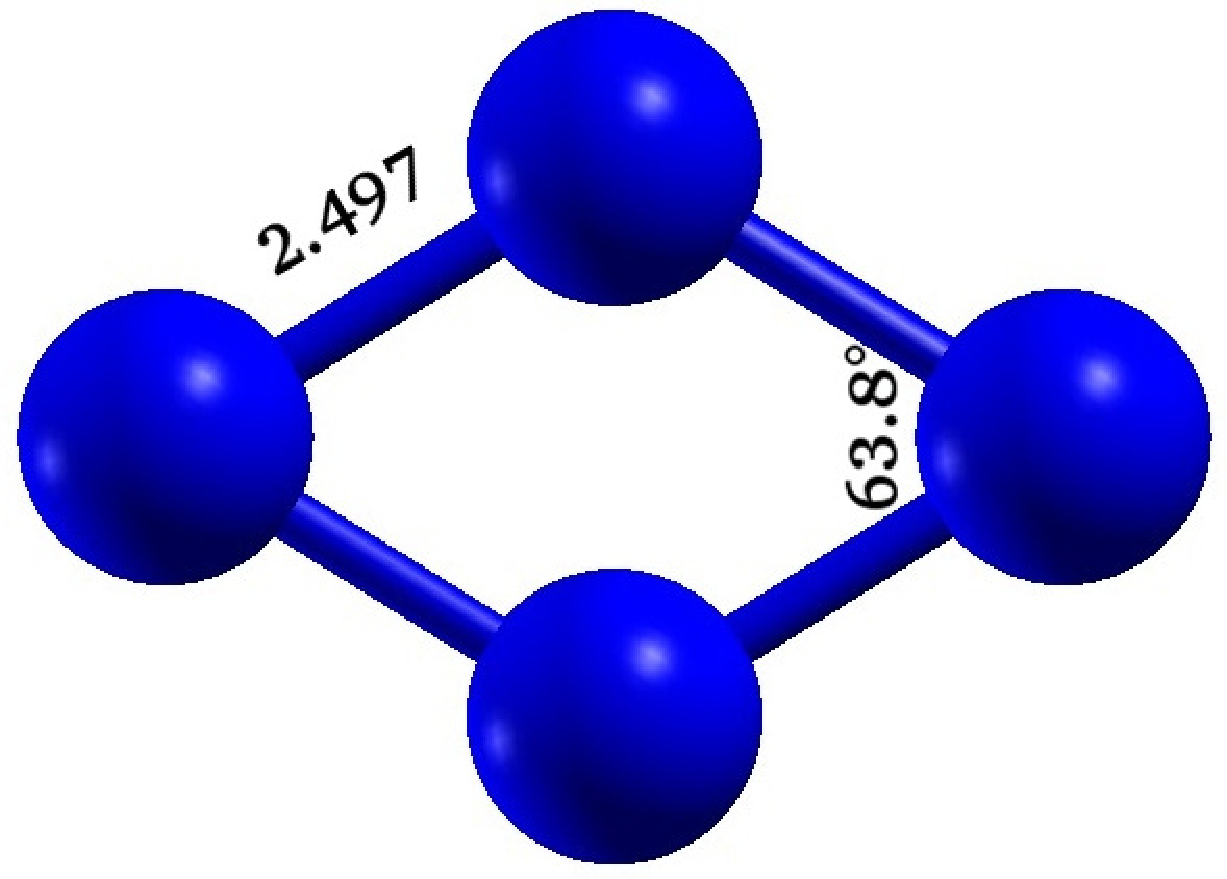}

}\hfill \subfloat[\textbf{Al$_{\boldsymbol{4}}$, D$_{\boldsymbol{4h}}$, $\boldsymbol{^{3}B{}_{3u}}$}]{\includegraphics[width=0.124\paperwidth]{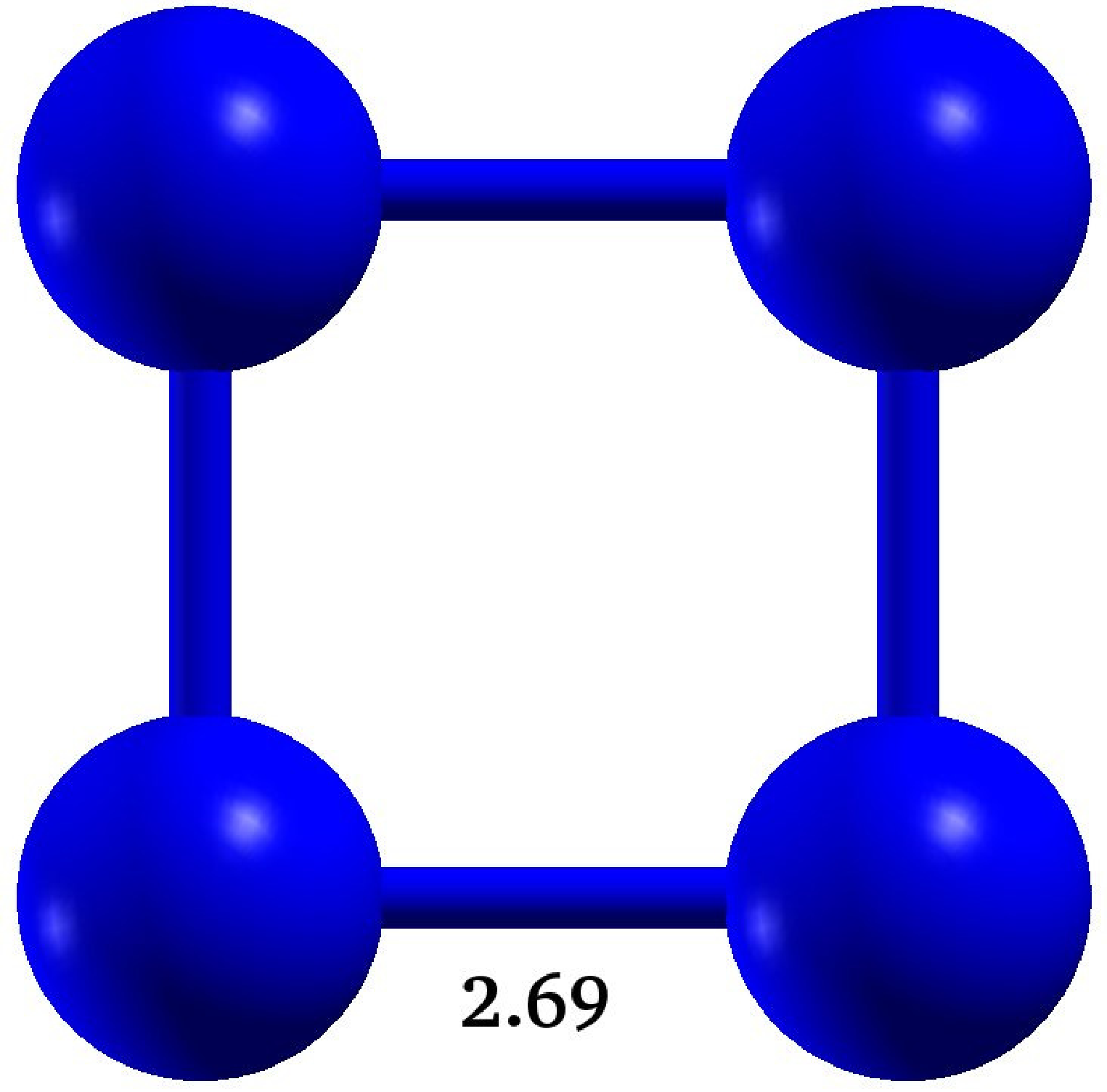}

}\hfill \subfloat[\textbf{Al$_{\boldsymbol{5}}$, C$_{\boldsymbol{2v}}$, $\boldsymbol{^{2}A{}_{1}}$}]{\includegraphics[width=0.2\paperwidth]{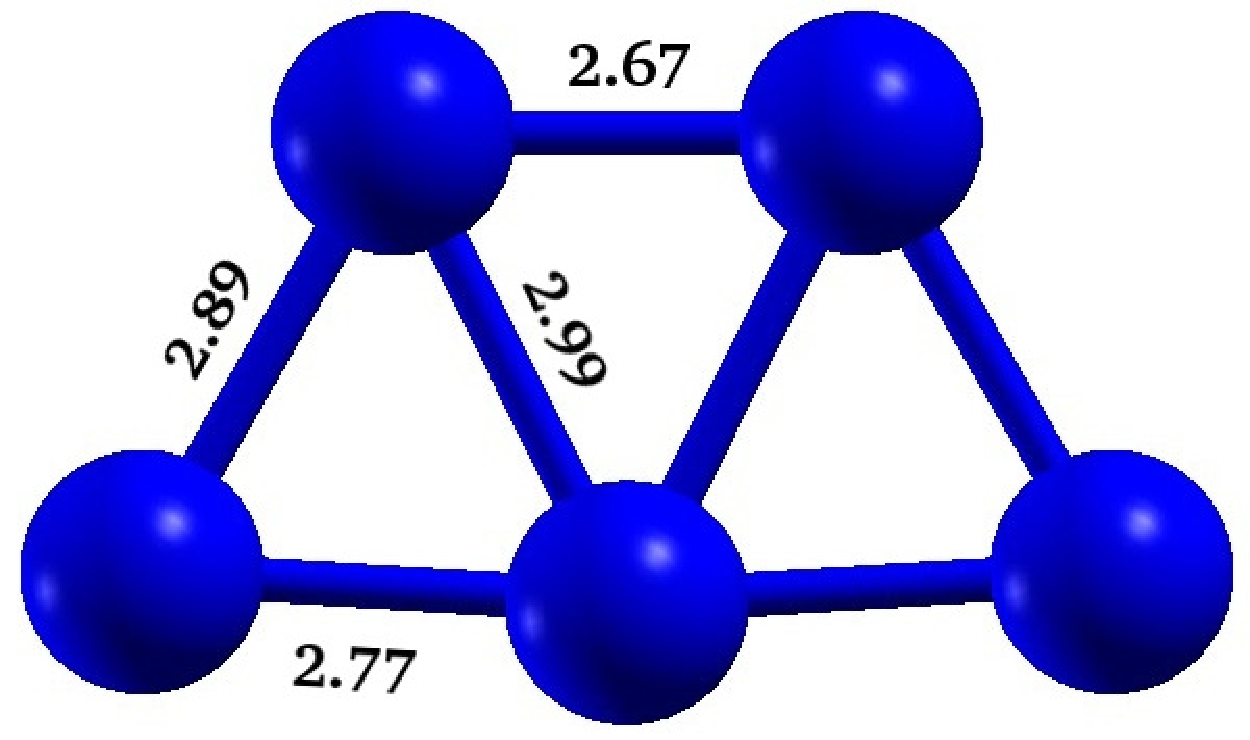}\label{subfig:geom-pentagon}

}\hfill \subfloat[\textbf{Al$_{\boldsymbol{5}}$, C$_{\boldsymbol{4v}}$,} $\boldsymbol{^{2}A{}_{1}}$]{\includegraphics[width=0.12\paperwidth]{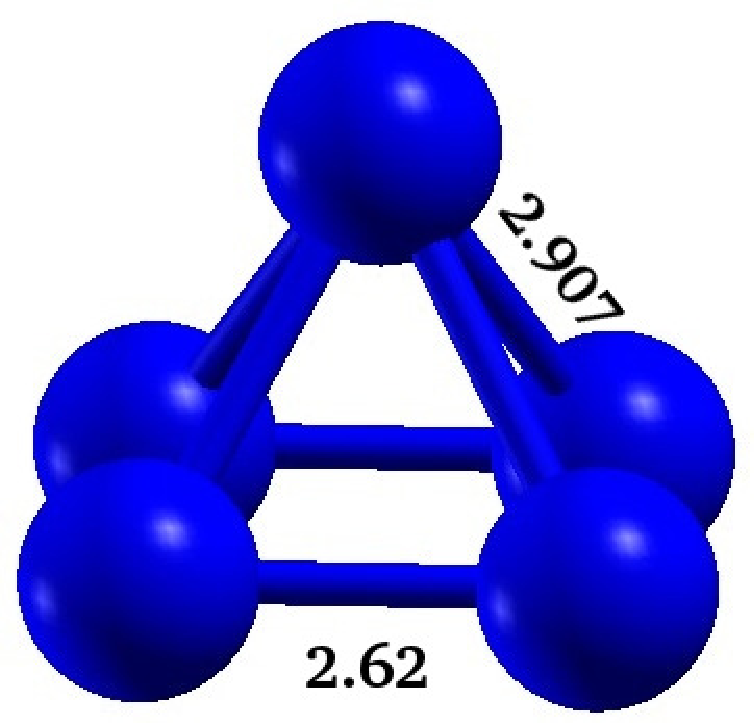}\label{fig:geom-pyra   }

} \vspace{0.2cm}

\caption{\label{fig:geometry}Geometry optimized structures of aluminum clusters with point group symmetry and the electronic
ground state at the CCSD level. All numbers are in $\textrm{\AA}$ unit. }
\end{figure*}

We repeated the optimization for singlet and triplet
systems on even numbered electron systems to look for the true ground state geometry. Similarly, for odd
numbered electron systems, doublet and quartet multiplicities were
considered in the geometry optimization. To initiate the optimization,
raw geometries, reported by Rao and Jena, based on density functional
method were used \cite{rao_jena_ele_struct_al}. Figure \ref{fig:geometry}
shows the final optimized geometries of the isomers studied in this
paper.

Using these optimized geometries, correlated calculations were performed
using multireference singles doubles configuration interaction (MRSDCI)
method for both ground state and excited states \cite{meld}. This
method considers a large number of singly- and doubly- substituted
configurations from a large number of reference configurations, and,
is well suited for both ground and excited states calculations. It
takes into account the electron correlations which are inadequately
represented in single reference \emph{ab initio} methods. These ground-
and excited-state wavefunctions are further used to calculate the
transition dipole moment matrix elements, which in turn, are utilized
to compute linear optical absorption spectrum assuming a Lorentzian
line shape.

Various wave functions of the excited states contributing to the peaks
in the spectrum obtained using a low-level CI calculations were analyzed, and even bigger MRSDCI calculations
were performed by including more references, if needed. The criteria of choosing
a reference configuration in the calculation was based upon the magnitude of the
corresponding coefficients in the CI wave function of the excited
states contributing to a peak in the spectrum. This process was repeated
until the spectrum converges within acceptable tolerance and all
the configurations which contribute to various excited states were
included. The typical total number of configurations considered in the calculations of various
isomers is given in Table \ref{tab:energies}. We have extensively
used such approach in performing large-scale correlated calculations
of linear optical absorption spectra of conjugated polymers \cite{mrsd_jcp_09,mrsd_prb_02,mrsd_prb_05,mrsd_prb_07},
and atomic clusters \cite{smallboron, borozene-sahu}.

The CI method is computationally very expensive,
mainly, because the number of determinants to be considered increases
exponentially with the number of electrons, and the number of molecular
orbitals. Calculations on bigger clusters are prohibitive under such
circumstances, and are very time consuming even for the clusters considered
here. Point group symmetries (D$_{2h}$, and its subgroups) were taken into
account, thereby making calculations for each symmetry subspace independent
of each other. The core of the aluminum atom was frozen from excitations,
keeping only three valence electrons active per atom. Also an upper limit on
the number of virtual orbitals was imposed, to restrict very high energy
excitations. The effect of these approximations on the computed photoabsorption
spectra has been studied carefully, and is presented in the next section.

\section{\label{sec:RESULTS-AND-DISCUSSION}Results and Discussion}

In this section, first we present a systematic study of the convergence
of our results and various approximations used. In the latter part,
we discuss the results of our calculations on various clusters.

\subsection{Convergence of calculations}

In this section we discuss the convergence of photoabsorption calculations
with respect to the choice of the basis set, and the size of the active
orbital space.

\subsubsection{Choice of basis set}

 In the literature several optimized basis sets are available for specific 
purposes, such as ground state optimization, excited state calculations etc. 
We have reported a systematic basis set dependence of photoabsorption
of boron cluster \cite{smallboron}. Similarly, here we have checked
the dependence of photoabsorption spectrum of aluminum dimer on basis
sets used,\cite{emsl_bas1,emsl_bas2} as shown in Fig. \ref{fig:basis-study}.
The 6-311 type Gaussian contracted basis sets are known to be good
for ground state calculations. The correlation consistent (CC) basis
sets, namely, CC-polarized valence double-zeta and CC-polarized valence
triple zeta (cc-pVTZ) give a good description of excited states of various systems.
The latter is found to be more sophisticated in describing the high
energy excitations, which were also confirmed using results of an
independent TDDFT calculation \cite{basis-set-core}. Therefore, in this work,  we have used 
the cc-pVTZ basis set for the optical absorption calculations.

\begin{figure}

\includegraphics[width=8cm]{FIG10.eps}\vspace{0.2cm}

\caption{\label{fig:basis-study}Optical absorption in Al$_{2}$
calculated using various Gaussian contracted basis sets.}

\end{figure}

\subsubsection{Orbital truncation scheme}

\vspace{0.5cm}

\begin{figure}

\includegraphics[width=8cm]{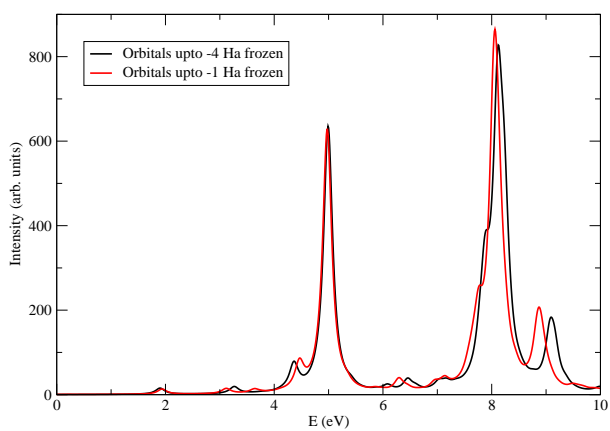}\vspace{0.2cm}

\caption{\label{fig:Core-Study}The effect of freezing the core
orbitals of aluminum atoms on optical absorption spectrum of Al$_{2}$.
It renders little effect on optical absorption spectrum, with significant
reduction in the computational cost.}

\end{figure}

With respect to the total number of orbitals N in the system, the
computational time in configuration interaction calculations scales
as $\approx$ N$^{6}$. Therefore, such calculations become intractable
for moderately sized systems, such as those considered here. So, in
order to make those calculations possible, the lowest lying molecular orbitals
are constrained to be doubly occupied in all the configurations, implying
that no virtual excitation can occur from those orbitals. It reduces
the size of the CI Hamiltonian matrix drastically. In fact, this approach
is recommended in quantum chemical calculations, because the basis
sets used are not optimized to incorporate the correlations in core
electrons \cite{szabo_book}. The effect of this approximation on the spectrum is as
shown in Fig. \ref{fig:Core-Study}. Since, calculations with all
electrons in active orbitals were unfeasible, we have frozen occupied
orbitals upto -4 Hartree of energy for the purpose of demonstration.
The effect of freezing the core is negligibly small in the low energy
regime, but shows disagreement in the higher energy range. However,
for very high energy excitations, photodissociation may occur, hence
absorption spectra at those energies will cease to have meaning. Thus,
the advantage of freezing the core subdues this issue. Therefore,
in all the calculations presented here, we have frozen the chemical core. 

\begin{figure}

\includegraphics[width=8cm]{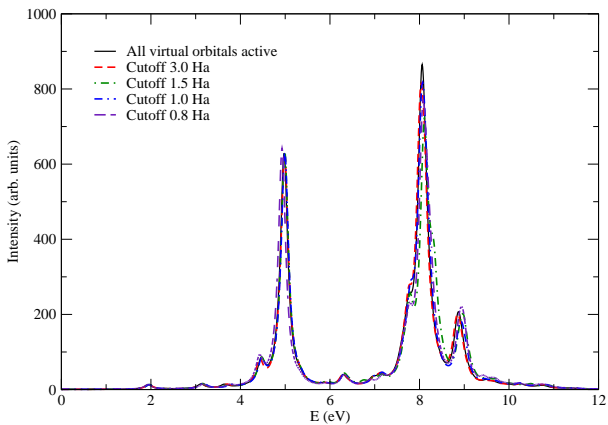}\vspace{0.2cm}

\caption{\label{fig:nref-study}The effect of the number of
active orbitals (N$_{act}$) on the optical absorption spectrum of
Al$_{2}$. Until N$_{act}$=46, the optical spectrum does not exhibit
any significant change. It corresponds to 1.0 Hartree ($\approx27.2$
eV) virtual orbital energy. }

\end{figure}

Not only occupied, but high energy virtual (unoccupied) orbitals can
also be removed from the calculations to make them tractable. In this
case the high lying orbitals are constrained to be unoccupied in all
the configurations. This move is justifiable, because it is unlikely
that electrons would prefer partial filling of high energy orbitals
in an attempt to avoid other electrons. However, this will only be
applicable if the orbitals are sufficiently high in energy. Fig. \ref{fig:nref-study}
shows the effect of removing orbitals having more than the specified
energy. From the figure it is clear that photoabsorption spectra exhibits
no difference at all up to 1 Hartree cutoff on virtual orbitals. Below
0.8 Ha cutoff, the spectra start deviating from each other. Hence,
we have ignored the virtual orbitals having energy more than 1 Ha. 
% \FloatBarrier

\subsubsection{Size of the CI expansion}

\begin{table*}
\small
\begin{centering}
\caption{The average number of total configurations (N$_{total}$) involved in MRSDCI calculations,
ground state (GS) energies (in Hartree) at the MRSDCI level, relative
energies and correlation energies (in eV) of various isomers of aluminum
clusters.\label{tab:energies}}

\par\end{centering}
\centering{}
% \begin{ruledtabular} %
\begin{tabular}{cccccc}
\hline
 Cluster  & Isomer  & N$_{total}$  & GS energy & Relative & \multicolumn{1}{c}{Correlation energy\textit{$^{c}$}}\tabularnewline
 &  &   & (Ha) & energy (eV) & \multicolumn{1}{c}{per atom(eV)}\tabularnewline
\hline 
Al$_{2}$  & Linear-I  & 445716 & -483.9138882 & 0.00 & 1.69\tabularnewline
          & Linear-II  & 326696 & -483.9115660 & 0.06 & 1.87\tabularnewline
 &  &  &  &  & \tabularnewline
Al$_{3}$  & Equilateral triangular  & 1917948 & -725.9053663 & 0.00 & 2.38\tabularnewline
 & Isosceles triangular &  1786700 & -725.8748996 & 0.83 & 2.36\tabularnewline
 & Linear &  1627016 & -725.8370397 & 1.85 & 2.16\tabularnewline
 &  &  &  &  & \tabularnewline
Al$_{4}$  & Rhombus  & 3460368 & -967.8665897 & 0.00 & 1.82\tabularnewline
 & Square &  1940116 & -967.8258673 & 1.11 & 1.80\tabularnewline
 &  &  &  &  & \tabularnewline
Al$_{5}$  & Pentagonal   & 3569914 & -1209.8114803 & 0.00 & 1.73\tabularnewline
 & Pyramidal  &  3825182 & -1209.7836568 & 0.76 & 1.77\tabularnewline
\hline
\end{tabular}
\end{table*}
\nofootnote{\textit{$^{c}$The difference in Hartree-Fock energy and MRSDCI correlated energy of the ground state.}}

In the multi-reference CI method, the size of
the Hamiltonian matrix increases exponentially with the number of molecular
orbitals in the system. Also, accurate correlated results can only
be obtained if sufficient number of reference configurations are included
in the calculations. In our calculations, we have included those configurations
which are dominant in the wave functions of excited states for a given
absorption peak. Also, for ground state calculations, we included
configurations until the total energy converges within a predefined
tolerance. Table \ref{tab:energies} shows the average number of total configurations involved  % text reference conf deleted
in the CI calculations of various isomers. For a given isomer, the
average is calculated across different irreducible representations
needed in these symmetry adapted calculations of the ground and various
excited states. For the simplest cluster, the total configurations
are about half a million and for the biggest cluster considered here,
it is around four million for each symmetry subspace of Al$_{5}$.
The superiority of our calculations can also be judged from the correlation
energy defined here (\emph{cf.} Table \ref{tab:energies}), which
is the difference in the total energy of a system at the MRSDCI level
and the Hartree-Fock level. The correlation energy per atom seems
to be quite high for all the clusters, making our calculations stand
out among other electronic structure calculations, especially single
reference DFT based calculations.

\subsection{Calculated photoabsorption spectra of various clusters}

In this section, we describe the photoabsorption spectra of various
isomers of the aluminum clusters studied. Plots of various molecular orbitals involved are presented 
in the Electronic Supporting Information (ESI).$\dagger$\cite{supplementary-material}

\subsubsection{Al$_2$}
Aluminum dimer is the most widely studied cluster of aluminum, perhaps because the nature of its ground state was a matter of debate for a long time. For example, in an early 
emission based experiment  Ginter \emph{et al.}~\cite{ginter_al2_apj_63} concluded that ground state of Al$_2$ was of symmetry $^3\Sigma^-_u$, while in a low-temperature absorption based experiment Douglas \emph{et al.}~\cite{absorption_spectra_al_dimer_jpc} deduced 
that the ground state of the system was of $^1\Sigma^-_g$. In other words, even the 
spin multiplicity of the cluster was measured to be different in different experiments. 
Theoreticians, on the other hand, were unanimous in predicting the spin multiplicity of the  ground state to be of triplet type, however, some predicted  $^3\Pi_u$ to be the ground 
state,~\cite{langhoff_3piu, electronic_state_dimer_cpl, sunil_and_jordan_al2, tse-al2-al3-1988, dft_al_dimer_jmolstruct} while others predicted it to be of $^3\Sigma^-_g$ 
type.~\cite{al_dimer_jphysb, upton_jphys_chem} Perhaps, the reason behind this ambiguity, was that states $^3\Pi_u$ and $^3\Sigma^-_g$ are located extremely close to each other as discovered in several theoretical calculations.~\cite{langhoff_3piu, electronic_state_dimer_cpl, sunil_and_jordan_al2, tse-al2-al3-1988, dft_al_dimer_jmolstruct} However,  it has now been confirmed experimentally by Cai \emph{et al.}\cite{bondybey_expt_3piu_dimer} and Fu \emph{et al.}\cite{spectra_jet_cooled_dimer_jcp} that the Al$_{2}$ (\emph{cf}. Fig. \ref{fig:geometry}\subref{subfig:subfig-al2})
has $^{3}\Pi_{u}$ ground state, with the $^3\Sigma^-_g$ state being a metastable state located slightly above it.
 
In our calculations, the bond length obtained using geometry optimization at CCSD level was 2.72 $\textrm{\AA}$, with D$_{\infty h}$
point group symmetry. This is in very good agreement with available data, such as Martinez \emph{et al}. obtained 2.73 $\textrm{\AA}$
as dimer length using all electron calculations \cite{martinez_all_vs_core}, 2.71 $\textrm{\AA}$ \cite{jones_struct_bind_prl} and 2.75 $\textrm{\AA}$ 
\cite{langhoff_3piu} as bond lengths using DFT and configuration interaction methods, and 2.86 $\textrm{\AA}$ obtained using DFT with
generalized gradient approximation \cite{rao_jena_ele_struct_al}. The experimental bond length of aluminum dimer is 2.70 $\textrm{\AA}$ \cite{bondybey_expt_3piu_dimer}.
We also performed the geometry optimization for the metastable state $^{3}\Sigma_{g}^{-}$ mentioned above, and found the bond length to be 2.48 $\textrm{\AA}$ (\emph{cf}. Fig. \ref{fig:geometry}\subref{subfig:subfig-al2}). Using MRCI calculations Bauschlicher \emph{et al.} estimated that $^{3}\Sigma_{g}^{-}$ electronic state lies 0.02 eV above 
the $^{3}\Pi_{u}$ ground state.\cite{langhoff_3piu} Our calculations predict  this difference to be 
about 0.06 eV.

\begin{figure}
\includegraphics[width=8cm]{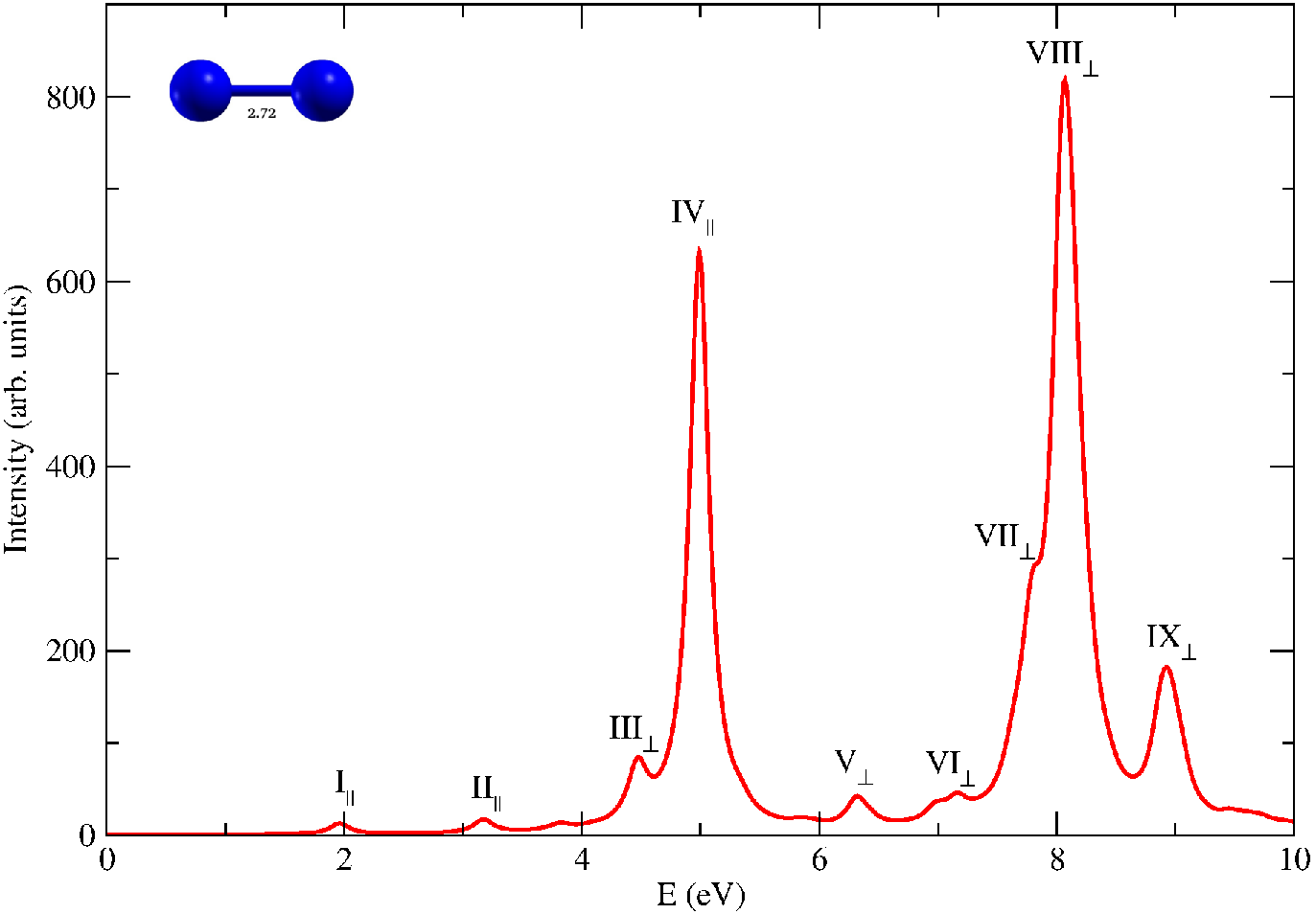}\vspace{-0.45cm} \\
\hspace{-0.1cm}\includegraphics[width=8.15cm]{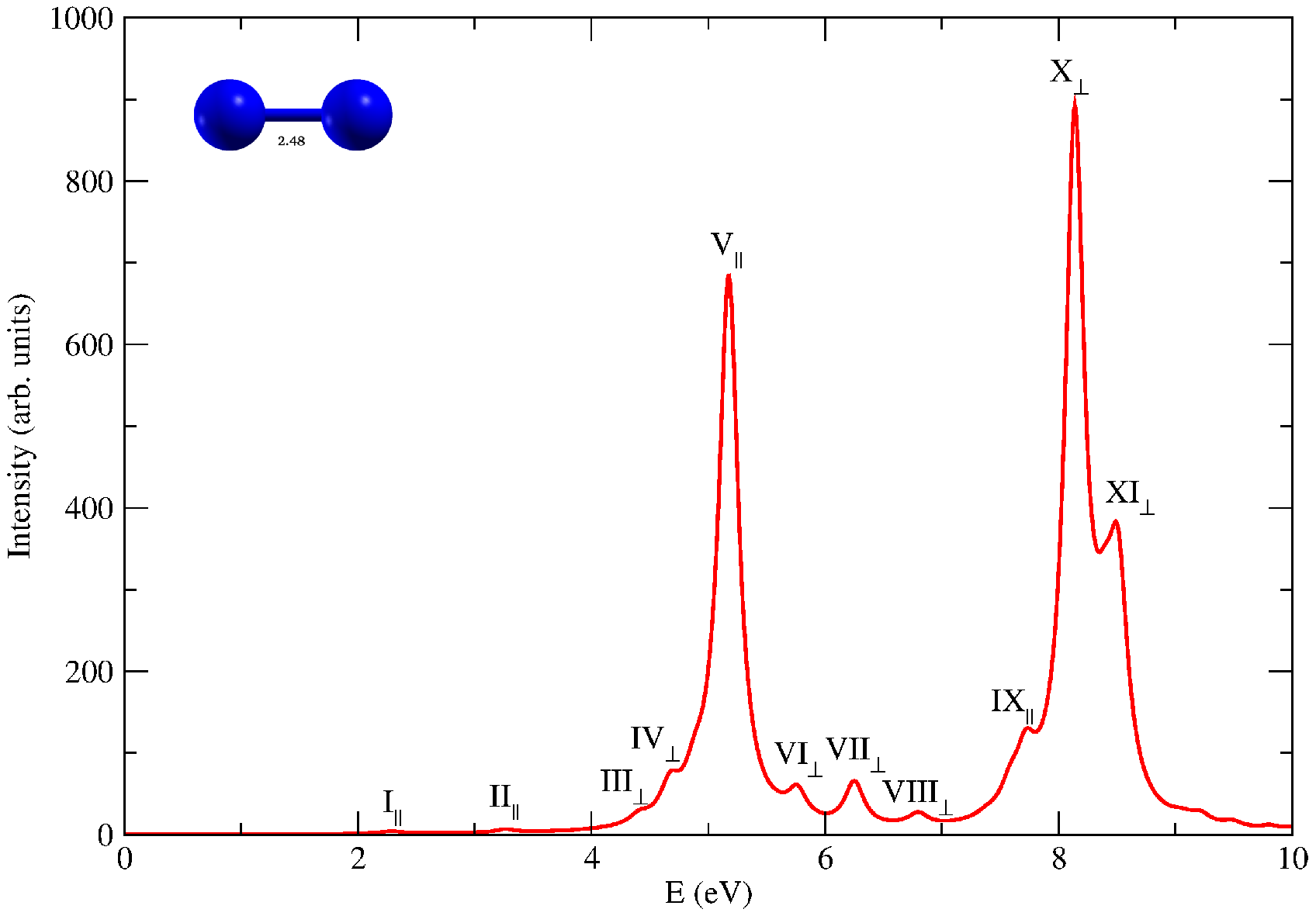}\vspace{0.2cm}
\caption{\label{fig:plot-al2-linear}The linear optical absorption
spectra of the global minimum Al$_{2}$ isomer ($^3\Pi_u$ state, top panel) and metastable isomer ($^3\Sigma^-_g$ state, bottom panel), 
calculated using the MRSDCI approach. The peaks corresponding to the light polarized along the molecular axis are
labeled with the subscript $\parallel$, while those polarized perpendicular
to it are denoted by the subscript $\perp$. For plotting the spectrum,
a uniform linewidth of 0.1 eV was used.}
\end{figure}

The many-particle wave function of Al$_{2}$ for the $^3\Pi_u$ ground state  consists of 
two degenerate singly occupied molecular orbitals (to be denoted by $H_{1}$ and
$H_{2}$, henceforth), because it is a spin triplet system. Similarly,
the configurations involving excitations from occupied molecular orbitals
to the unoccupied orbitals, form excited state wave functions. The
computed photoabsorption spectra of Al$_{2}$, as shown in Fig. \ref{fig:plot-al2-linear},
is characterized by weaker absorption at lower energies and couple
of intense peaks at higher energies. The many-particle wave functions
of excited states contributing to the peaks are presented in Table I of supporting information \cite{supplementary-material}.  
The spectrum starts with a small absorption
peak (I$_{\parallel}$) at around 2 eV, characterized by $H_{2}\rightarrow L+1$ and
light polarized along the direction of axis of the dimer. It is followed
by a couple of small intensity peaks (II$_{\parallel}$, III$_{\perp}$), until a dominant absorption (IV$_{\parallel}$)
is seen at 5 eV. This is characterized by $H_{1}\rightarrow L+3$.
Another dominant peak (VIII$_{\perp}$) is observed at 8 eV having $H-2\rightarrow L$
as dominant configuration, with absorption due to light polarized
perpendicular to the axis of the dimer. 

The optical absorption spectrum of metastable dimer in the $^3\Sigma^-_g$ state (\emph{cf.} Fig. \ref{fig:plot-al2-linear}) is 
also characterized by small absorption peaks in the lower energy range.
Also, all peaks of the spectrum appear blue-shifted as compared to that of stable isomer. The peak (I$_{\parallel}$) at 2.29 eV is characterized by 
$H - 1 \rightarrow L$, while two major peaks  at 5.17 eV (V$_{\parallel}$) and  8.13 eV (X$_{\perp}$) are characterized by $H -1 \rightarrow L$ configuration due to 
light polarized along the direction of axis of dimer and $H -1 \rightarrow L+1$ due to transversely polarized absorption 
respectively.

Douglas \emph{et al.}\cite{absorption_spectra_al_dimer_jpc} obtained the low-energy optical absorption in the cryogenic krypton matrix. 
The major peaks in this experimental absorption spectrum at 1.77 eV and 3.13 eV can be associated with our 
results of 1.96 eV and 3.17 eV respectively. Although, our calculation overestimates the location of the first peak by 
about 11\%, the agreement between theory and experiment is excellent for the second peak, giving us confidence about the quality our calculations. Furthermore, the computed spectrum for the 
metastable state $^3\Sigma^-_g$ of Al$_2$ (\emph{cf.} Fig. \ref{fig:plot-al2-linear}) has no 
peaks close to 
those observed in the experiments, implies  that measured optical absorption occurs in the    
$^{3}\Pi_{u}$ state of the system, confirming that the ground state has   $^{3}\Pi_{u}$ symmetry.

% In our calculations, we have observed a  $^{3}\Pi_{u}$ $\rightarrow$ $^{3}\Sigma_{g}^{-}$ 
% transition at 0.22 eV, but it has negligible intensity.

Our spectrum differs from the one obtained with the time-dependent
local density approximation (TDLDA) method \cite{kanhere_prb_optical_al} 
in both the intensity and the number of peaks. 
However, we agree with TDLDA\cite{kanhere_prb_optical_al} in 
predicting two major peaks at  5 eV (IV$_{\parallel}$) and  8 eV (VIII$_{\perp}$). 
Unlike our calculations, the number of peaks is much more in TDLDA results and the spectrum is almost
continuous. Peaks located in our calculations at 3.2 eV (II$_{\parallel}$) and 6.3 eV (V$_{\perp}$) are also observed in the 
TDLDA spectrum of the dimer,\cite{kanhere_prb_optical_al} except for the fact that in our calculations both the peaks are relatively minor, while the TDLDA calculation predicts the 6.3 eV peak to be fairly intense.

\subsubsection{Al$_3$}

Among the possible isomers of aluminum cluster Al$_{3}$, the equilateral triangular isomer is found to be the most stable. We have considered
three isomers of Al$_{3}$, namely, equilateral triangle, isosceles triangle, and a linear chain. The most stable isomer has $D{}_{3h}$ point
group symmetry, and $^{2}A_{1}^{'}$ electronic state. The optimized bond length 2.57 $\textrm{\AA}$, is in good agreement with reported
theoretical values 2.61 $\textrm{\AA}$ \cite{rao_jena_ele_struct_al}, 
2.62 $\textrm{\AA}$ \cite{upton_elec_struct_small_al}, 2.56 $\textrm{\AA}$ \cite{martinez_all_vs_core}, 2.54  $\textrm{\AA}$ \cite{low-lying-al3-baeck-jcp}
 and 2.52 $\textrm{\AA}$ \cite{sachdev_dft_al_small,truhler_theory_validation_jpcb}.
The doublet ground state is also confirmed with the results of magnetic deflection experiments \cite{cox-al3-experiment}. 

The next isomer, which lies 0.83 eV higher in energy, is the isosceles triangular
isomer. The optimized geometry has 2.59 $\textrm{\AA}$, 2.59 $\textrm{\AA}$
and 2.99 $\textrm{\AA}$ as sides of triangle, with a quartet ground state ($^{4}A_{2}$). Our results are in agreement
with other theoretical results \cite{upton_elec_struct_small_al,martinez_all_vs_core,jones_simul_anneal_al}.

Linear Al$_{3}$ isomer again with quartet multiplicity is the next
low-lying isomer. The optimized bond length is 2.62 $\textrm{\AA}$.
This is in good agreement with few available reports \cite{martinez_all_vs_core,jones_simul_anneal_al,sachdev_dft_al_small}.

Li \emph{et al.} reported infrared optical absoption in Al$_{3}$ in inert-gas matrices at low temperature.\cite{infrared_spectra_boron_aluminum_cpl} 
Another experimental study of optical absorption in isosceles triangular isomer was performed by Fu \emph{et al.} using 
jet cooled aluminum clusters.\cite{spectra_jet_cooled_trimer_jcp,spectra_trimer_ijms} 

\begin{figure}
\hspace{-0.1cm}\includegraphics[width=8.12cm]{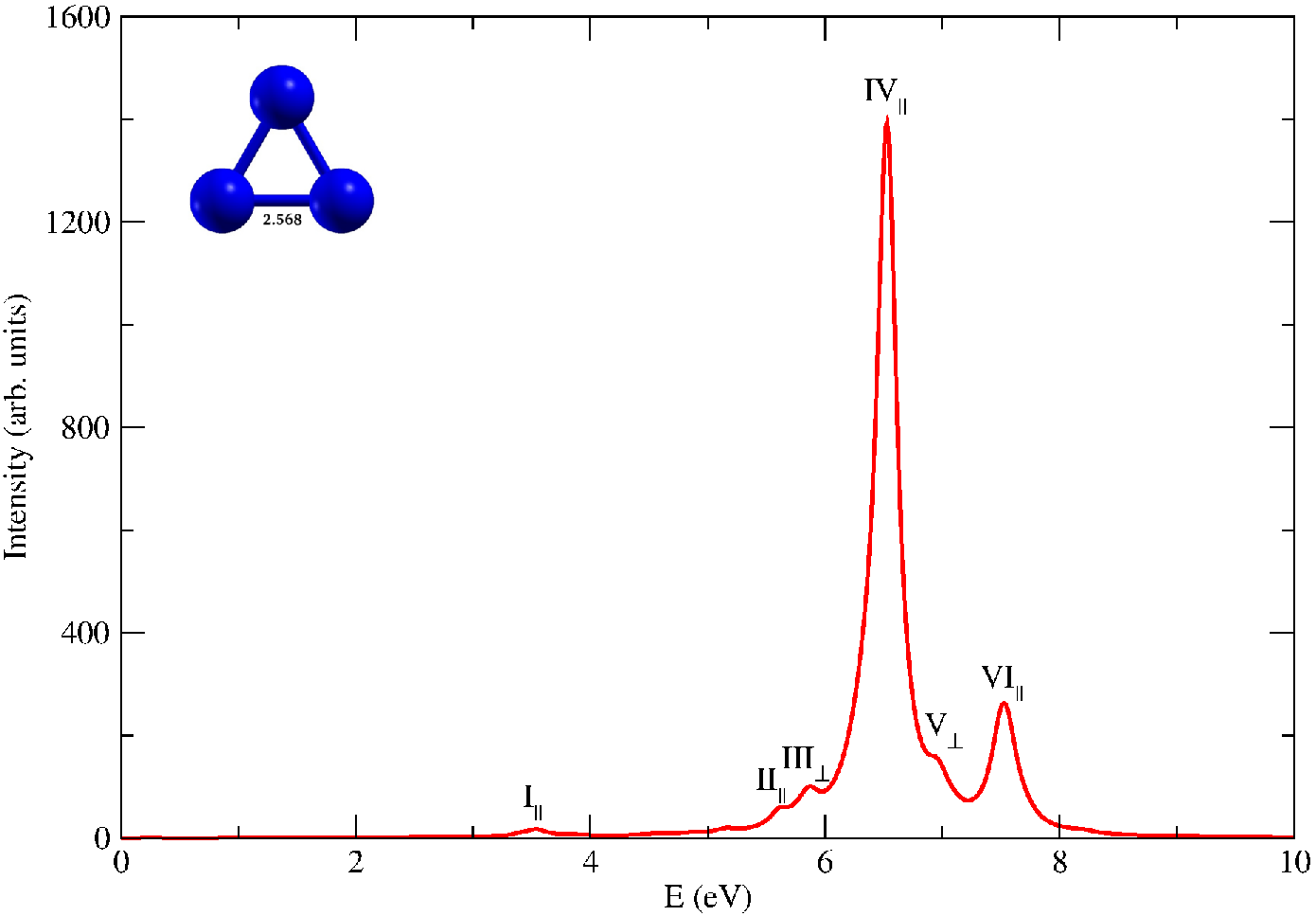} \vspace{-0.45cm} \\
\includegraphics[width=8cm]{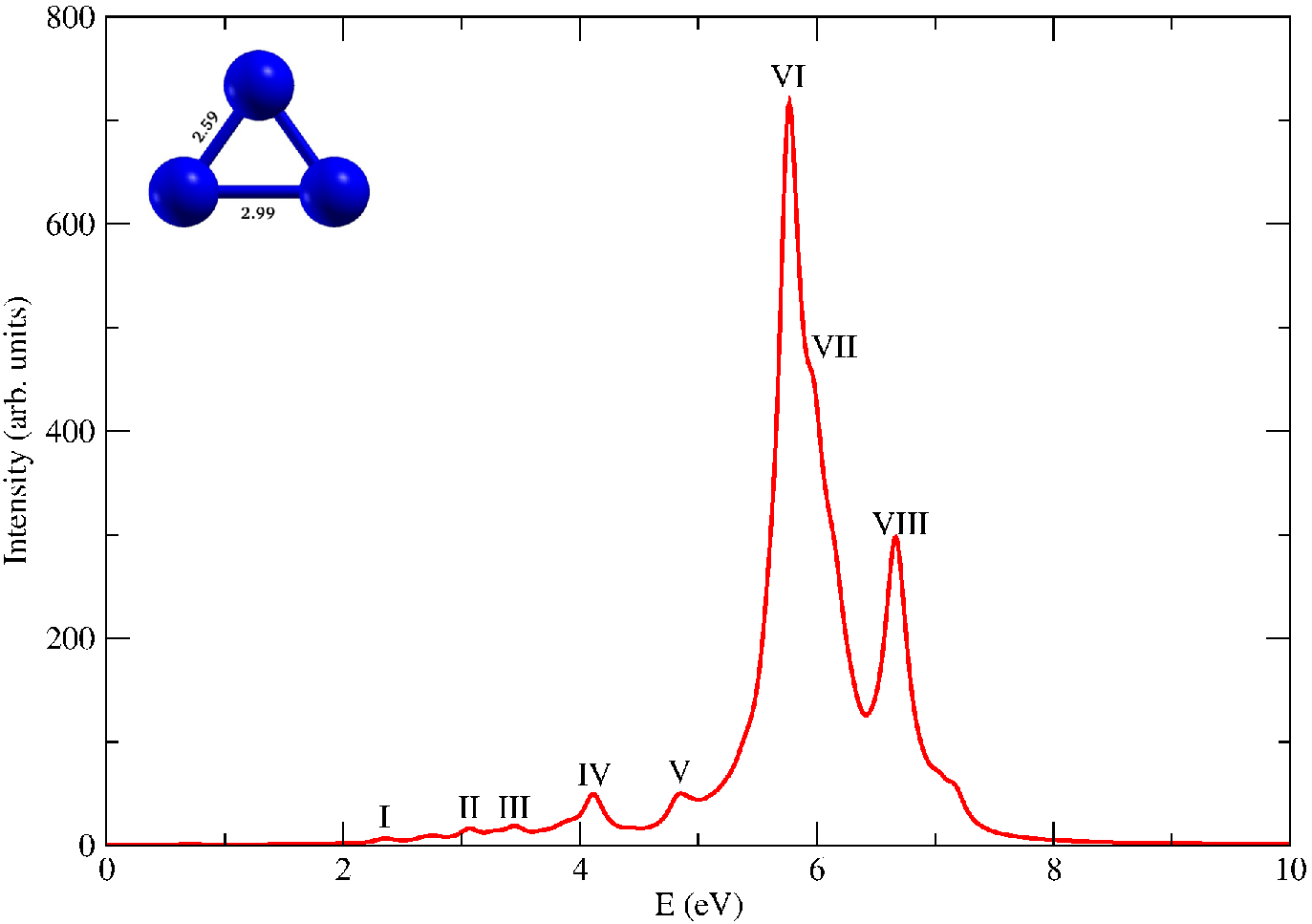} \vspace{-0.45cm} \\
\includegraphics[width=8cm]{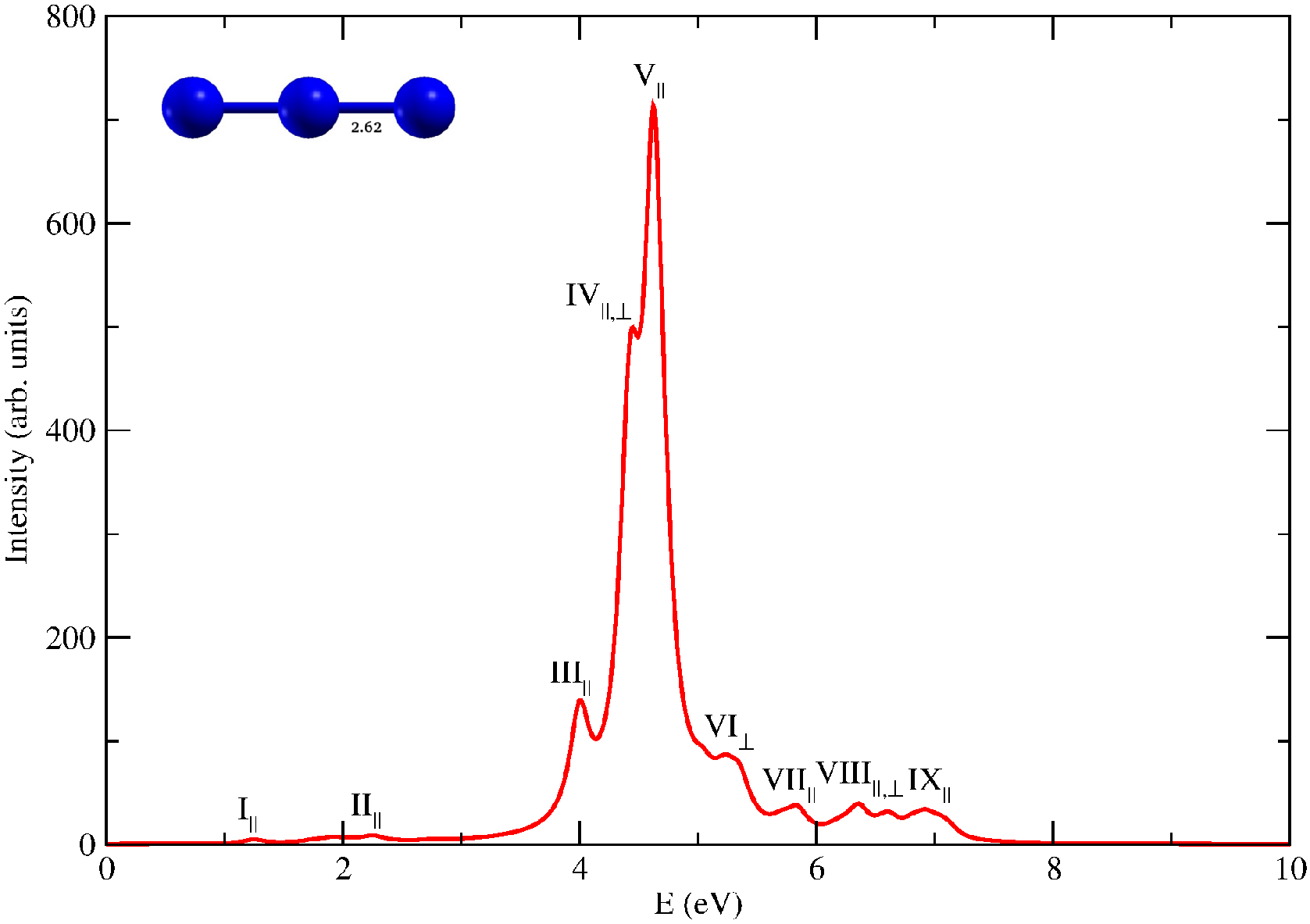} \vspace{0.2cm}
\caption{\label{fig:plot-al3-equil}The linear optical absorption
spectra of Al$_{3}$ equilateral triangle isomer, isosceles isomer, and linear isomer calculated using
the MRSDCI approach. The peaks corresponding to the light polarized
along the molecular plane are labeled with the subscript $\parallel$,
while those polarized perpendicular to it are denoted by the subscript
$\perp$. All peaks in the spectrum of isosceles isomer correspond to the light polarized along the molecular plane.
Rest of the information is same as given in the caption of Fig. \ref{fig:plot-al2-linear}}
\end{figure}

The photoabsorption spectra of these isomers are presented in Fig. \ref{fig:plot-al3-equil}. %, \ref{fig:plot-al3-iso} and \ref{fig:plot-al3-lin}.
The corresponding many body wave functions of excited states corresponding
to various peaks are presented in Table III, IV and V of supporting information \cite{supplementary-material}. %Table \ref{Tab:table_al3_equil_tri},
%\ref{Tab:table_al3_iso_tri} and \ref{Tab:table_al3_lin_tri} respectively.
In the equilateral triangular isomer, most of the intensity is concentrated
at higher energies. The same is true for the isosceles triangular
isomer. However, the spectrum of isosceles triangular isomer appears
slightly red shifted with respect to the equilateral counterpart.
Along with this shift, there appears a split pair of peaks at 5.8
eV (VI and VII). This splitting of oscillator strengths is due to distortion
accompanied by symmetry breaking. The absorption spectrum of linear
isomer is altogether different with bulk of the oscillator strength
carried by peaks in the range 4 -- 5 eV, and, due to the polarization
of light absorbed parallel to the axis of the trimer.

\FloatBarrier
The optical absorption spectrum of equilateral triangular isomer consists
of very feeble low energy peaks at 3.5 eV (I$_{\parallel}$), 5.6 eV (II$_{\parallel}$) and 5.8 eV (III$_{\perp}$) characterized
by $H-3\rightarrow L+5$, a double excitation $H-2\rightarrow L+5;H-1\rightarrow L+5$,
and $H-3\rightarrow L+2$ respectively. The latter peak is due to
the light polarized perpendicular to the plane of the isomer. It is
followed by an intense peak (IV$_{\parallel}$) at around 6.5 eV with dominant contribution
from $H\rightarrow L+6$ and $H\rightarrow L+4$ configurations. A
semi-major peak (VI$_{\parallel}$) is observed at 7.5 eV characterized mainly due to
double excitations.

Two major peaks at 6.5 eV (IV$_{\parallel}$) and 7.5 eV (VI$_{\parallel}$) in the spectrum of Al$_{3}$ equilateral isomer, obtained in our calculations
are also found in the spectrum of TDLDA calculations, with the difference
that the latter does not have a smaller intensity in TDLDA \cite{kanhere_prb_optical_al}. Other
major peaks obtained by Deshpande \emph{et al.} \cite{kanhere_prb_optical_al}
in the spectrum of aluminum trimer are not observed, or have very small
intensity in our results.

As compared to the equilateral triangle spectra, the isosceles triangular
isomer with quartet spin multiplicity, exhibits several small intensity peaks (\emph{cf}. Fig. \ref{fig:plot-al3-equil})
in the low energy regime. The majority of contribution to peaks of
this spectrum comes from in-plane polarized transitions, with
negligible contribution from transverse polarized light. The spectrum
starts with a feeble peak (I$_{\parallel}$) at 2.4 eV with contribution from doubly-excited
configuration $H\rightarrow L+1;H-2\rightarrow L+2$. Although, no experimental absorption data is available for the doublet equilateral triangle isomer, 
Fu \emph{et al.}\cite{spectra_trimer_ijms, spectra_jet_cooled_trimer_jcp} managed to measure the absorption of the isosceles triangle isomer, and observed this peak to be around 2.5 eV. Thus, this excellent agreement between the experiment 
and our theoretical calculations for isosceles triangle  isomer with quartet
 spin multiplicity, further strengthens our belief in the quality of our calculations.
One of the dominant contribution to the oscillator strength comes from two closely-lying
peaks (VI$_{\parallel}$ and VII$_{\parallel}$) at 5.8 eV. The wave functions of excited states corresponding
to this peak show a strong mixing of doubly-excited configurations,
such as $H-3\rightarrow L+1;H-2\rightarrow L$ and $H-2\rightarrow L+1;H-4\rightarrow L$.
The peak (VIII$_{\parallel}$) at 6.7 eV shows absorption mainly due to $H\rightarrow L+10$.

Linear trimer of aluminum cluster also shows low activity in the low
energy range. Very feeble peaks are observed at 1.2 eV (I$_{\parallel}$) and 2.3 eV (II$_{\parallel}$),
both characterized by $H-3\rightarrow H-2$. This configuration also
contributes to the semi-major peak (III$_{\parallel}$) at 4 eV along with $H-4\rightarrow H$.
Two closely lying peaks at 4.3 eV (IV$_{\parallel,\perp}$) and 4.6 eV (V$_{\parallel}$) carry the bulk of the
oscillator strength. Major contribution to the former comes from $H-1\rightarrow L+2$
along with $H-3\rightarrow H-2$ being dominant in both the peaks.
Again, as expected, the absorption due to light polarized along the
trimer contributes substantially to the spectrum.

It is obvious from the spectra presented above that the location of the most 
intense absorption is quite sensitive to the structure, and thus can be 
used to distinguish between the three isomers.

\subsubsection{Al$_4$}

\begin{figure}
\includegraphics[width=8cm]{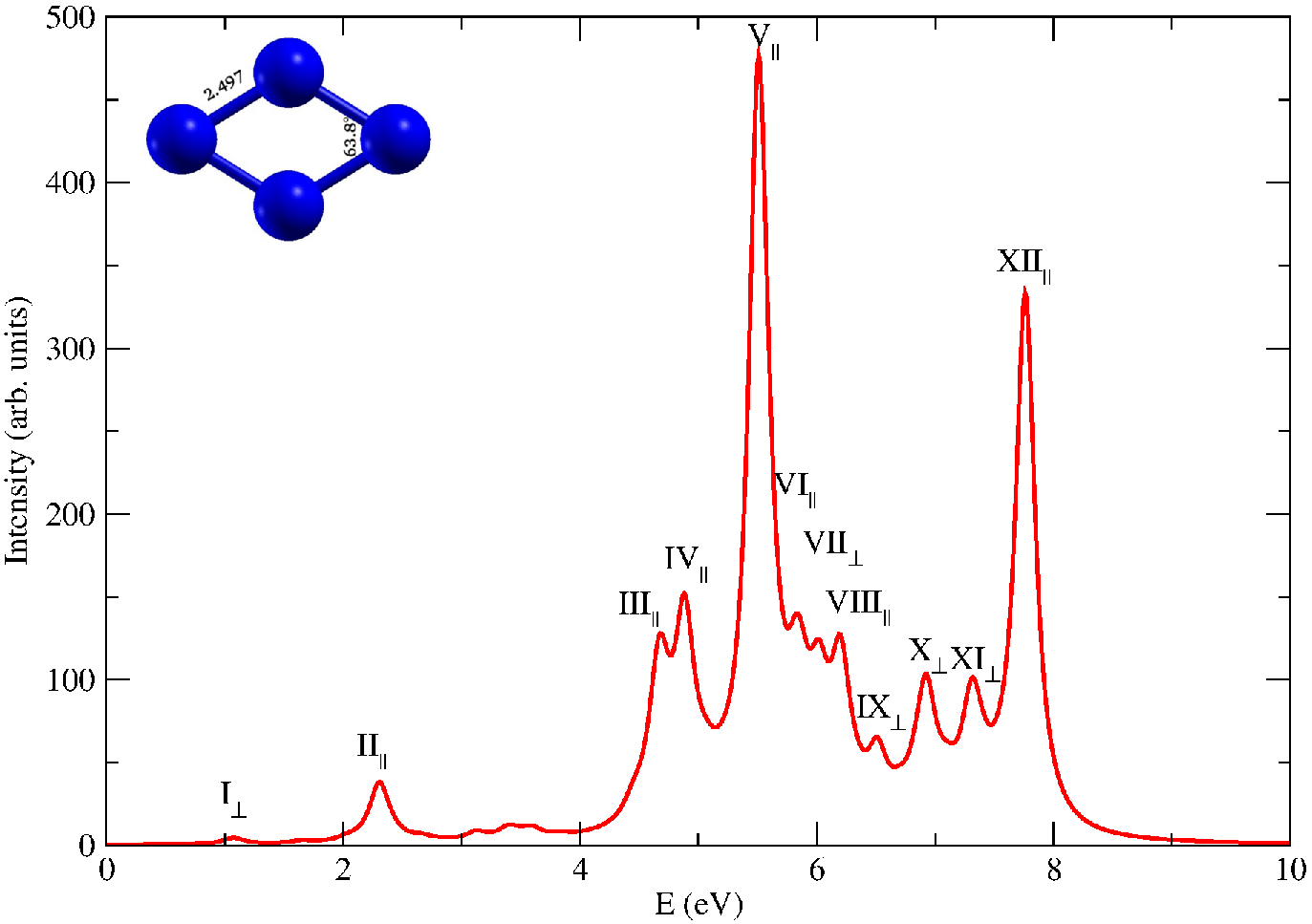} \vspace{-0.45cm} \\
\hspace{-0.19cm}\includegraphics[width=8.11cm]{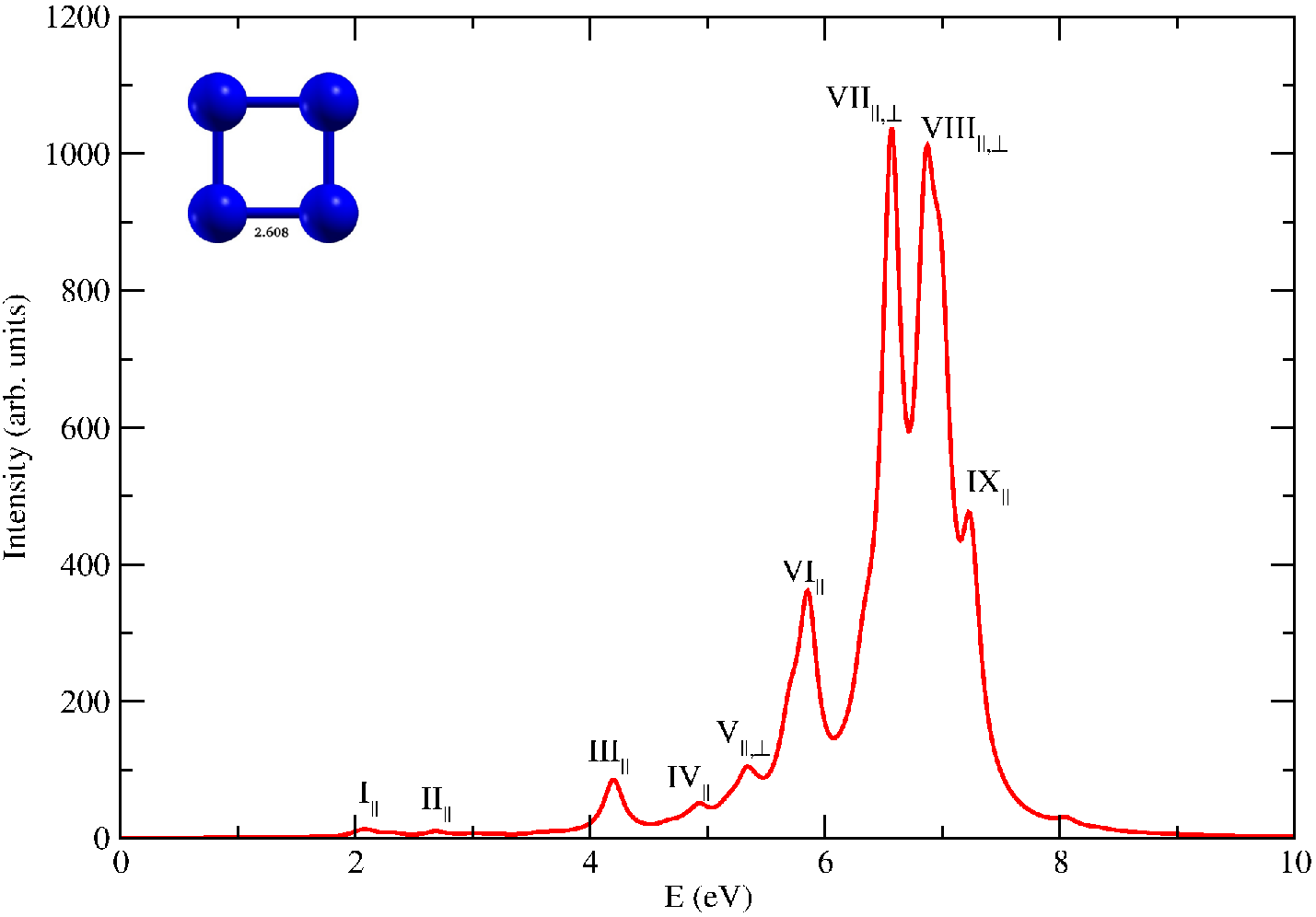}\vspace{0.2cm}
\caption{\label{fig:plot-al4-rho}The linear optical absorption
spectra of rhombus and square isomers of Al$_{4}$, calculated using the MRSDCI approach.
Rest of the information is same as given in the caption of Fig. \ref{fig:plot-al3-equil}.
}
\end{figure}

Tetramer of aluminum cluster has many low lying isomers due to its
flat potential energy curves. Among them, rhombus structure is the
most stable with $^{3}B_{2g}$ electronic ground state. Our optimized
bond length for rhombus structure is 2.50 $\textrm{\AA}$ and $63.8^\circ$
as the acute angle. This is to be compared with corresponding reported
values of 2.56 $\textrm{\AA}$ and $69.3^\circ$ reported by Martinez
\emph{et al.} \cite{martinez_all_vs_core}, 2.51 $\textrm{\AA}$
and $56.5^\circ$ computed by Jones \cite{jones_struct_bind_prl},
2.55 $\textrm{\AA}$ and $67.6^\circ$ obtained by Schultz \emph{et
al} \cite{truhler_theory_validation_jpcb}. We note that bond lengths
are in good agreement but bond angles appear to vary a bit.

The other isomer studied here is a square shaped tetramer
with optimized bond length of 2.69 $\textrm{\AA}$. The electronic
ground state of this $D_{4h}$ symmetric cluster is\textbf{ $^{3}B{}_{3u}$}\emph{.}
This optimized geometry is in accord with 2.69 $\textrm{\AA}$ reported
by Martinez \emph{et al}. \cite{martinez_all_vs_core}, however,
it is somewhat bigger than 2.57 $\textrm{\AA}$ calculated by Yang
\emph{et al.} \cite{sachdev_dft_al_small} and 2.61 $\textrm{\AA}$
obtained by Jones \cite{jones_simul_anneal_al}.

For planar clusters, like rhombus and square shaped Al$_{4}$, two
types of optical absorptions are possible: (a) planar -- those
polarized in the plane of the cluster, and (b) transverse -- the ones
polarized perpendicular to that plane. The many-particle wave functions
of excited states contributing to the peaks are presented in Table VI and VII of supporting information \cite{supplementary-material}. 
 The onset of optical absorption in rhombus isomer occurs at around 1 eV (I$_{\perp}$) with transversely polarized
absorption characterized by $H_{1}\rightarrow L+1$. It is followed
by an in-plane polarized absorption peak (II$_{\parallel}$) at 2.3 eV with dominant contribution
from $H-2\rightarrow H_{1}$. Several closely lying peaks are observed
in a small energy range of 4.5 -- 8 eV. Peaks split from each other
are seen in this range confirming that after shell closure, in perturbed
droplet model, Jahn Teller distortion causes symmetry breaking usually
associated with split absorption peaks. The most intense peak (V$_{\parallel}$) is observed
at 5.5 eV characterized by $H-3\rightarrow L+4$.

The absorption spectrum of square shaped isomer begins with a couple
of low in-plane polarized absorption peaks at 2.1 eV (I$_{\parallel}$) and 2.7
eV (II$_{\parallel}$) characterized by $H-1\rightarrow L$ and $H_{2}\rightarrow L+1$
respectively. The peak at 4.2 (III$_{\parallel}$) and 4.9 eV (IV$_{\parallel}$) have $H-2\rightarrow L$
and $H_{1}\rightarrow L+2$ as respective dominant configurations.
A major peak (VI$_{\parallel}$) at 5.85 eV is observed with absorption due to in-plane
polarization having $H-2\rightarrow L+2$ and a double excitation
$H_{1}\rightarrow L+2;H-2\rightarrow L+2$ as dominant configurations.
These configurations also make dominant contribution to the peak (VII$_{\parallel,\perp}$) at
6.5 eV. This peak along with one at 6.9 eV (VIII$_{\parallel,\perp}$) are two equally and most
intense peaks of the spectrum. The latter has additional contribution
from the double excitation $H_{1}\rightarrow L+1;H-2\rightarrow L$. A shoulder peak (IX$_{\parallel}$) is
observed at 7.2 eV.

The TDLDA spectrum \cite{kanhere_prb_optical_al} of aluminum rhombus tetramer differs from the one presented here.
Peaks labeled III to XII in our calculated spectrum are
also observed in the TDLDA results\cite{kanhere_prb_optical_al}, however, the relative 
intensities tend to disagree.  For example, the strongest absorption
peak of TDLDA calculations is located around 7.9 eV, while in our spectrum we obtain the 
second most intense peak at that location.    The highest absorption
peak (V$_{\parallel}$) in our calculations is at 5.5 eV, while TDLDA does report a strong
peak at the same energy\cite{kanhere_prb_optical_al}, it is not the highest of the spectrum. 

Our calculations also reveal a strong structure-property relationship as far as the location 
of the most intense peak in the absorption spectra of the two isomers is considered, a 
feature which can be utilized in their optical detection.

\subsubsection{Al$_5$}

The lowest lying pentagonal isomer of aluminum has $C_{2v}$ symmetry
and has an electronic ground state of $^{2}A_{1}$. The bond lengths
are as shown in Fig. \ref{fig:geometry}\subref{subfig:geom-pentagon}.
These are slightly bigger than those obtained by Rao and Jena \cite{rao_jena_ele_struct_al}
and Yang \emph{et al.} \cite{sachdev_dft_al_small} using the DFT approach.
Many other reports have confirmed that the planar pentagon is the
most stable isomer of $\textrm{Al}_{_{5}}$.

The other optimized structure of pentamer is perfect pyramid with
$C_{4v}$ symmetry and $^{2}A_{1}$ electronic ground state. This
lies 0.76 eV above the global minimum structure. This is the only
three dimensional structure studied in this paper for optical absorption.
The optimized geometry is consistent with those reported earlier by
Jones \cite{jones_simul_anneal_al}. However, it should be noted
that there exists many more similar or slightly distorted structure
lying equally close the the global minimum.

\begin{figure}
\includegraphics[width=8cm]{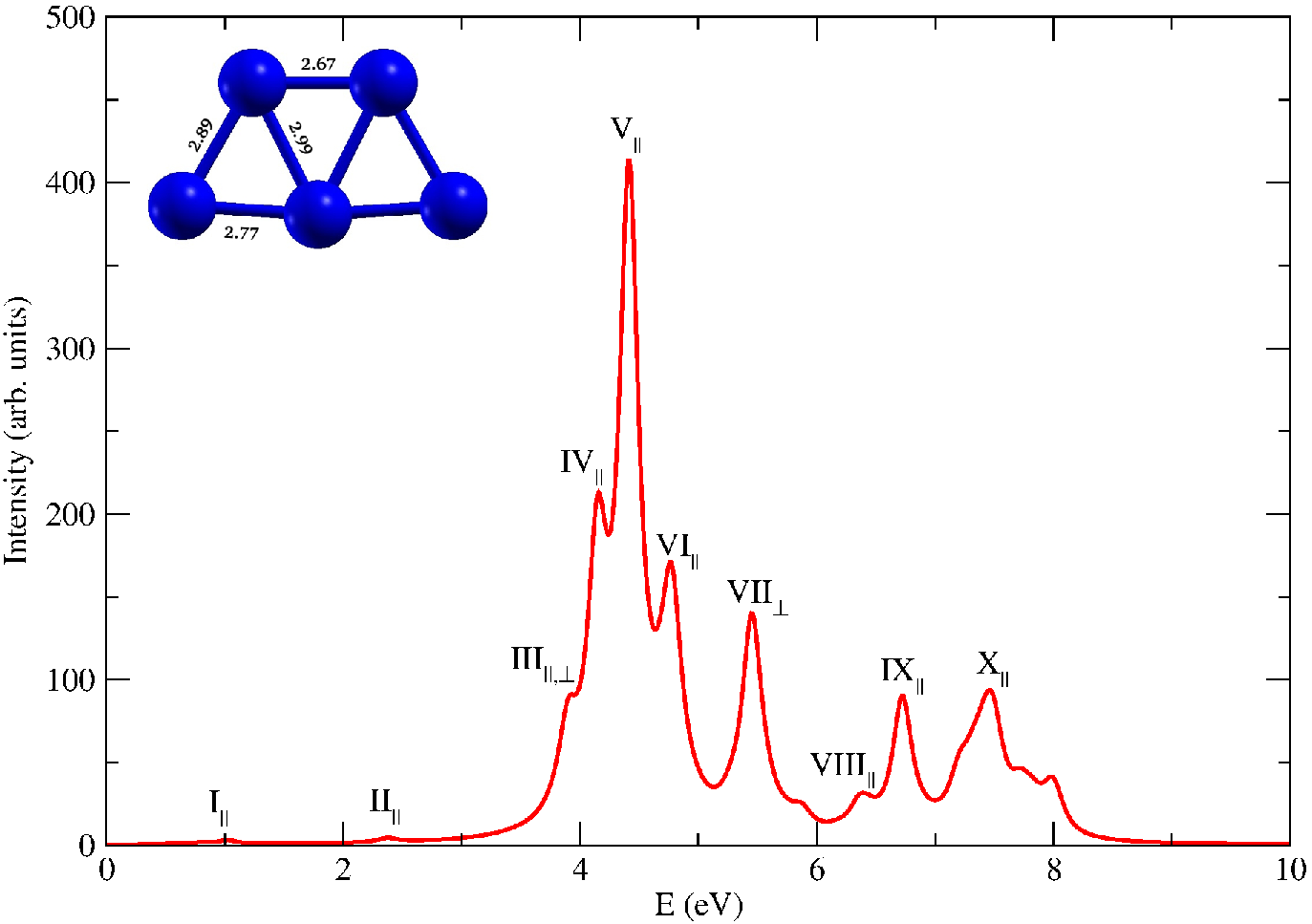} \vspace{-0.45cm} \\
\hspace{-0.13cm}\includegraphics[width=8cm]{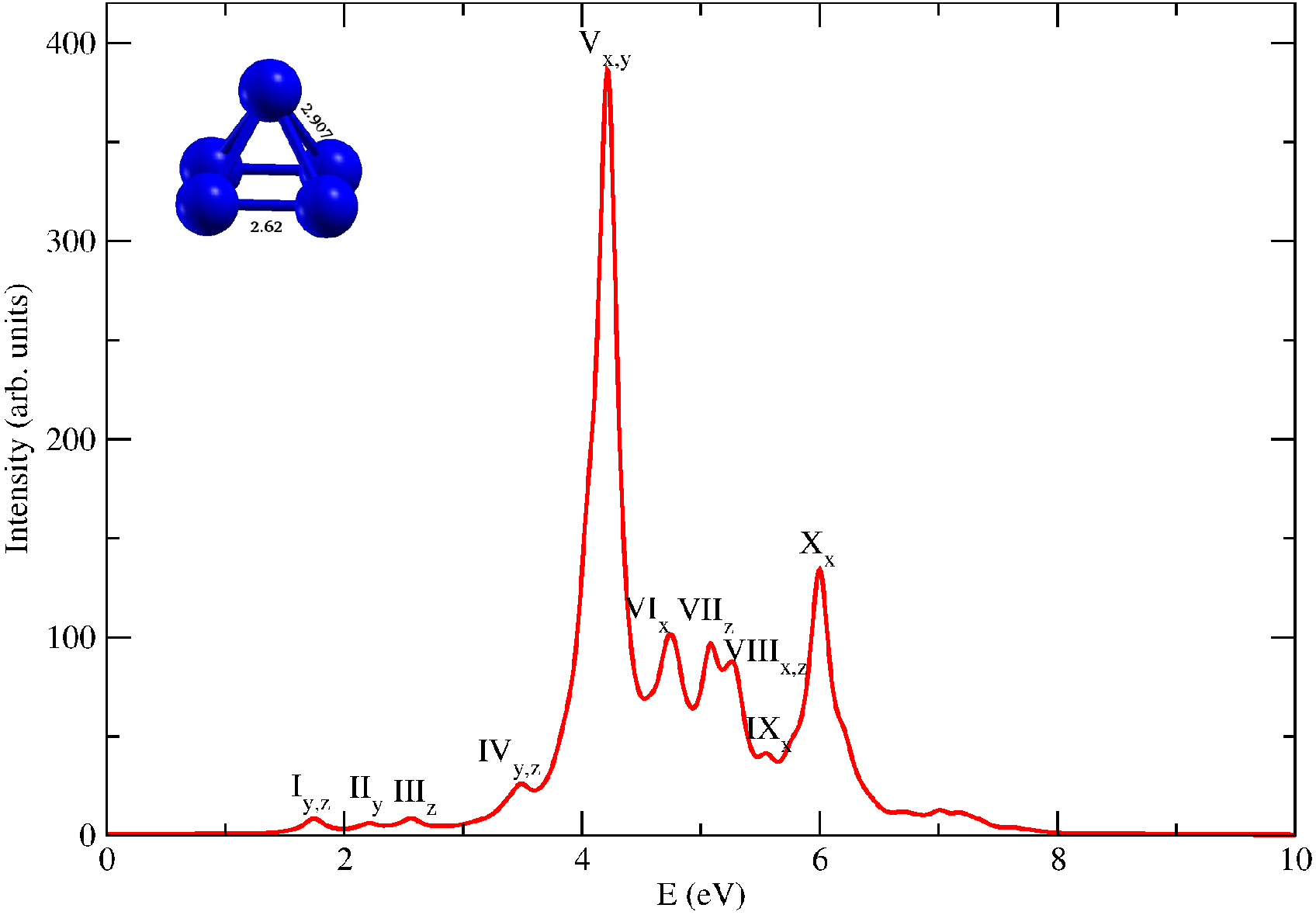}\vspace{0.2cm}
\caption{\label{fig:plot-al5-penta}The linear optical absorption
spectra of pentagonal and pyramidal Al$_{5}$, calculated using the MRSDCI approach.
 The peaks in the spectrum of pyramidal isomer corresponding to the light polarized along the Cartesian axes are labeled accordingly. Cartesian $xy$
  plane is assumed parallel to the base of the pyramid.
Rest of the information is same as given in the caption of Fig. \ref{fig:plot-al3-equil}.}
% The peaks corresponding to the light polarized along the molecular
% axis are labeled with the subscript $\parallel$, while those polarized
% perpendicular to it are denoted by the subscript $\perp$. For plotting
% the spectrum, a uniform linewidth of 0.1 eV was used.}
\end{figure}

The many-particle wave functions of excited states contributing to the 
peaks are presented in Table VIII and IX of supporting information \cite{supplementary-material}.  
The optical absorption spectrum of pentagonal Al$_{5}$ (Fig. \ref{fig:plot-al5-penta}) has few low
energy peaks followed by major absorption (V$_{\parallel}$) at 4.4 eV. It has dominant
contribution from $H-1\rightarrow L+5$ configuration. Pentagonal
isomer shows more optical absorption in the high energy range, with
peaks within regular intervals of energy.

Few feeble peaks occur in the low energy range in the optical absorption
of pyramidal isomer. The major absorption peak (V$_{x,y}$) at 4.2 eV is slightly
red-shifted as compared to the pentagonal counterpart. It is characterized
by $H-3\rightarrow L+2$. A peak (X$_{x}$) at 6 eV is seen in this absorption
spectrum having dominant contribution from $H\rightarrow L+13$, which
is missing in the spectrum of pentagon. These differences can lead
to identification of isomers produced experimentally.

In the range of spectrum studied in our calculations, the TDLDA calculated spectrum 
\cite{kanhere_prb_optical_al} of pentagonal isomer is found to be similar to the one presented here as far as the peak 
locations are concerned, albeit
the intensity profile differs at places. A small peak at 2.4 eV (II$_{\parallel}$) is observed in both the spectra, followed by 
peaks at 3.9 eV (III$_{\parallel,\perp}$), 4.2 eV (IV$_{\parallel}$) and 4.4 eV (V$_{\parallel}$).
These three peaks are also observed in TDLDA results with a little
bit of broadening. Again, the peak at 5.4 eV (VII$_{\perp}$) matches with each other
calculated from both the approaches. Peak found at 6.7 eV (IX$_{\perp}$) is also
observed in the TDLDA calculation\cite{kanhere_prb_optical_al}. Within the energy 
range studied here, the strongest peak position and intensity of this work is
in good agreement with that of its TDLDA counterpart\cite{kanhere_prb_optical_al}.

\section{\label{sec:CONCLUSIONS-AND-OUTLOOK}Conclusions and Outlook}

In this study, we have presented large-scale all-electron correlated
calculations of optical absorption spectra of several low-lying isomers
of aluminum clusters Al$_{n}$ (n=2--5), involving valence transitions. 
The present study does not take into account Rydberg transitions, 
which are more of atomic properties, than molecular ones. 
Both ground and excited
state calculations were performed at MRSDCI level, which take electron
correlations into account at a sophisticated level.  We have analyzed
the nature of low-lying excited states.  We see strong configuration mixing in various excited
states indicating plasmonic nature of excitations as per the criterion
suggested by Blanc \emph{et al.}\cite{plasmon}

 Isomers of a given cluster
show a distinct signature spectrum, indicating a strong structure-property relationship, which 
is usually found in small metal clusters. Such structure-property relationship exists for photoelectron spectroscopy as well, therefore, the optical absorption spectroscopy can be used as an alternative probe of the structures of clusters, and can be employed in experiments to distinguish
between different isomers of a cluster. The optical absorption spectra of few isomers of 
aluminum dimer and trimer are in very good agreement with the available experimental results.
Owing to the sophistication of our calculations,  our results can be used for benchmarking 
of the absorption spectra. Furthermore, our calculations demonstrate that the MRSDCI approach, within a first-principles formalism, can be used to perform sophisticated calculations of not just the ground state, but also of the excited states of metal clusters, in a 
numerically efficient manner. Moreover, by 
using more diffuse basis functions, one can also compute the Rydberg transitions, in case their 
description is warranted. 

Our results were found to be significantly different as compared to the 
TDLDA results\cite{kanhere_prb_optical_al}, for the clusters studied here.
This disagreement could be resolved by future optical absorption experiments performed on these clusters.

\section*{Acknowledgments} 
One of us (R.S.) would like to acknowledge the Council of Scientific and Industrial Research (CSIR), India, for 
research fellowship (09/087/(0600)2010-EMR-I). We also acknowledge CDAC, Pune for providing computational facility Param Yuva -II.
.
 \bibliography{smallal}
\newpage

\begin{center} 
{\large \bf Supplementary material for Large-scale first principles configuration interaction calculations of optical absorption in 
aluminum clusters}
\end{center}

In this document, we present the plots of the most important  molecular orbitals of  the
isomers of aluminum clusters considered in this work, and depicted in  Fig. 1 of the paper. 
Furthermore, we also present their ground and excited state CI wave functions, energies, 
and oscillator strengths corresponding to various peaks in their photo-absorption spectra 
discussed in section III of the paper.

\section{Molecular Orbitals of Aluminum Clusters}
 \begin{figure}[b]
 \includegraphics[width=7cm]{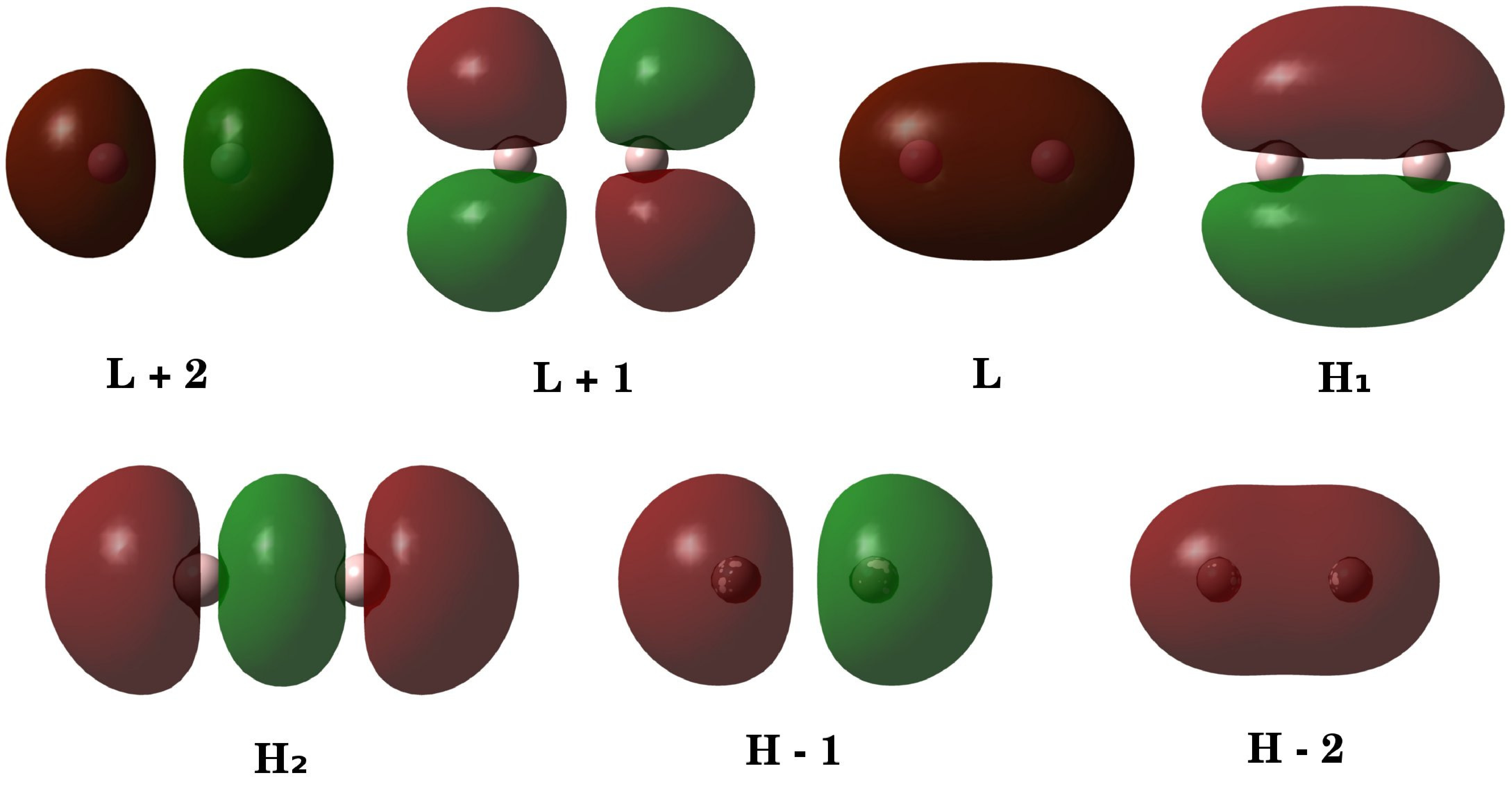}
 \caption{Molecular orbitals of aluminum dimer. $H$ and $L$
 stands for HOMO and LUMO respectively, and $H_{1}$ and $H_{2}$ are
 singly occupied degenerate molecular orbitals.}
 \end{figure}
 
 \begin{figure}
 \includegraphics[width=7cm]{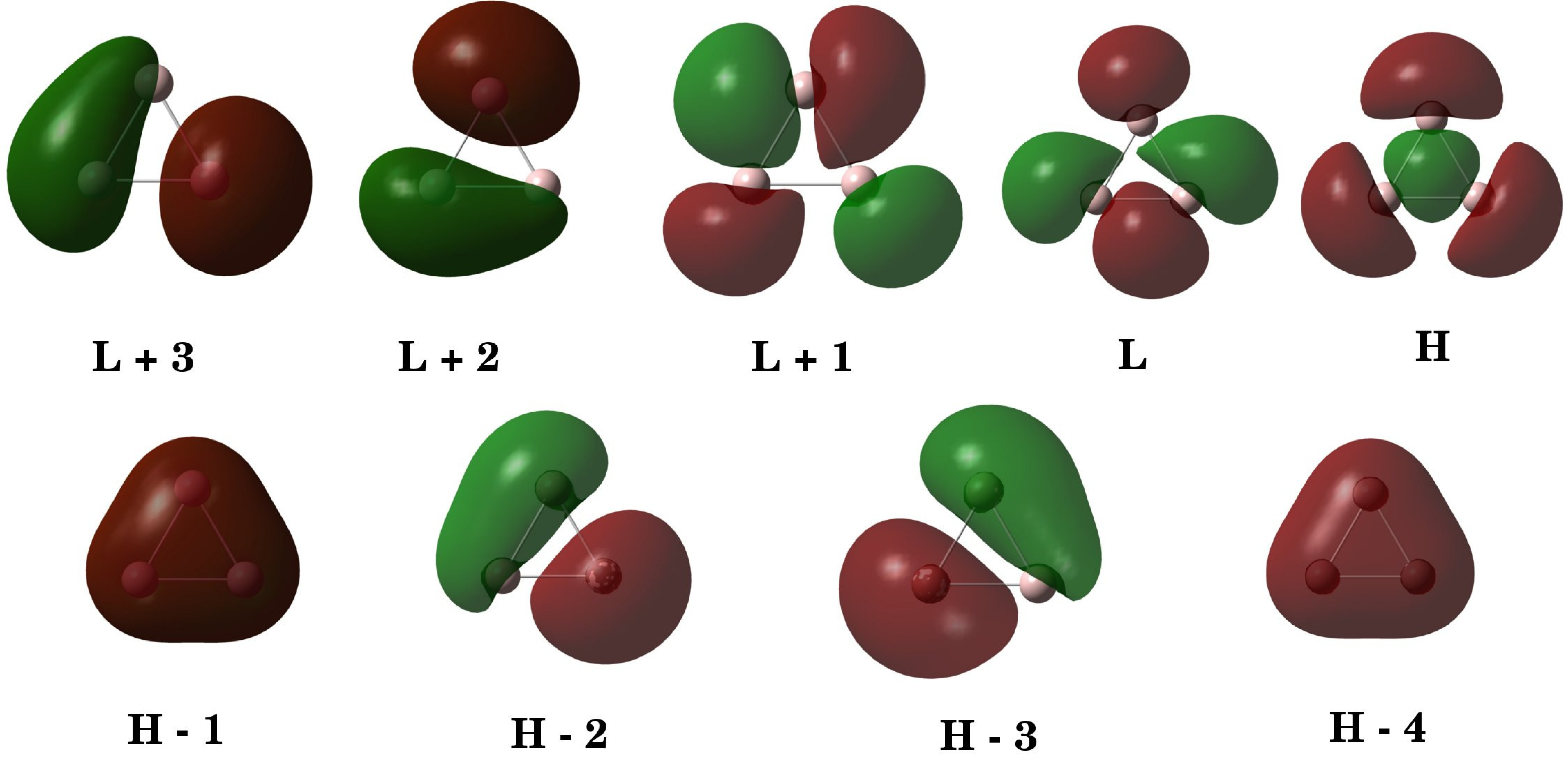} 
 \caption{Molecular orbitals of equilateral triangular aluminum
 trimer. $H$ and $L$ stands for HOMO and LUMO respectively. ($H-2$, $H-3$), ($L$, $L+1$) and ($L+2$, $L+3$) are degenerate pairs.}
 \end{figure}

 \begin{figure}
 \includegraphics[width=7cm]{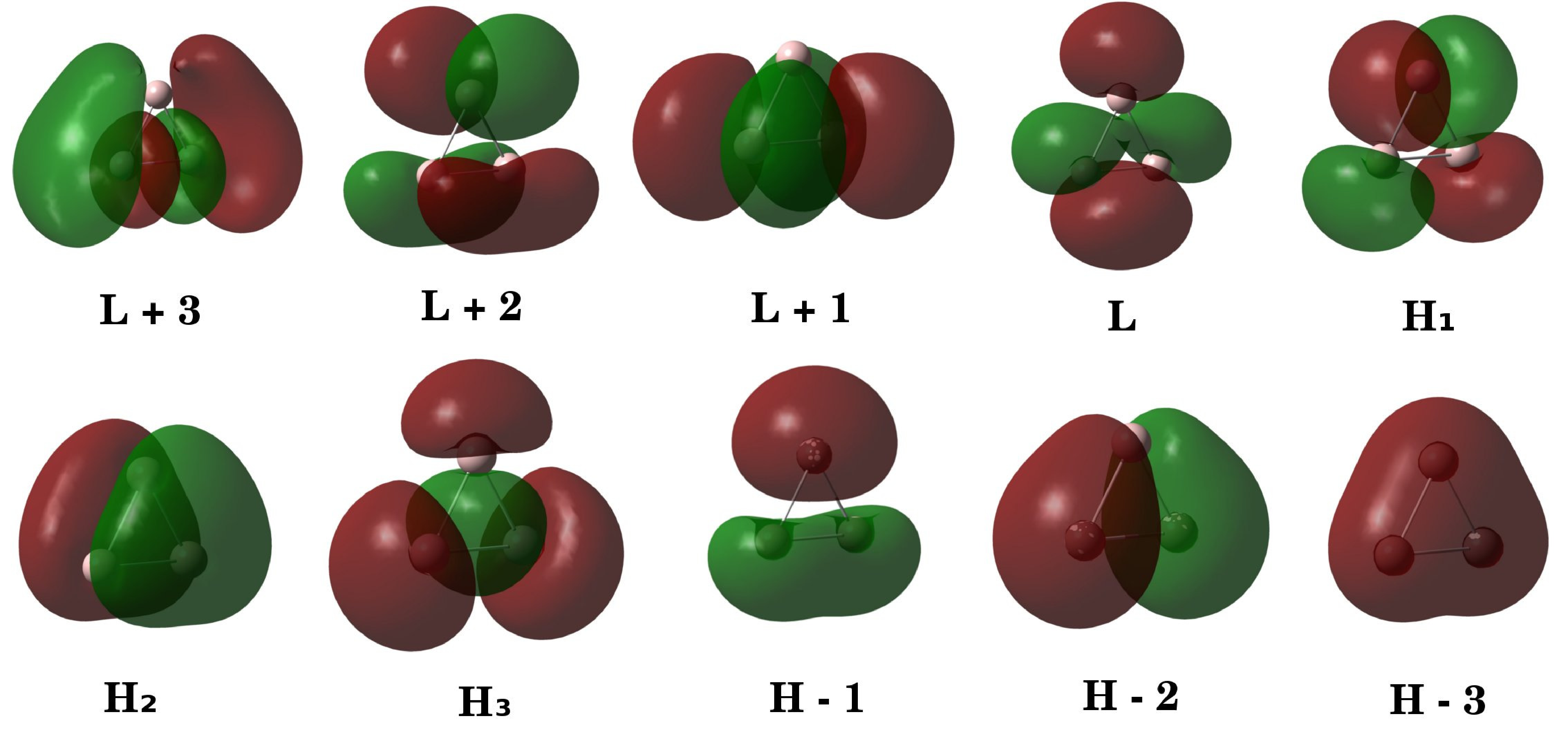}
 \caption{Molecular orbitals of isosceles triangular aluminum
  trimer. $H$ and $L$ stands for HOMO and LUMO respectively, and $H_{1}$,
 $H_{2}$, and $H_{3}$ are singly occupied molecular orbitals.}
 \end{figure}

 \begin{figure}
 \includegraphics[width=7cm]{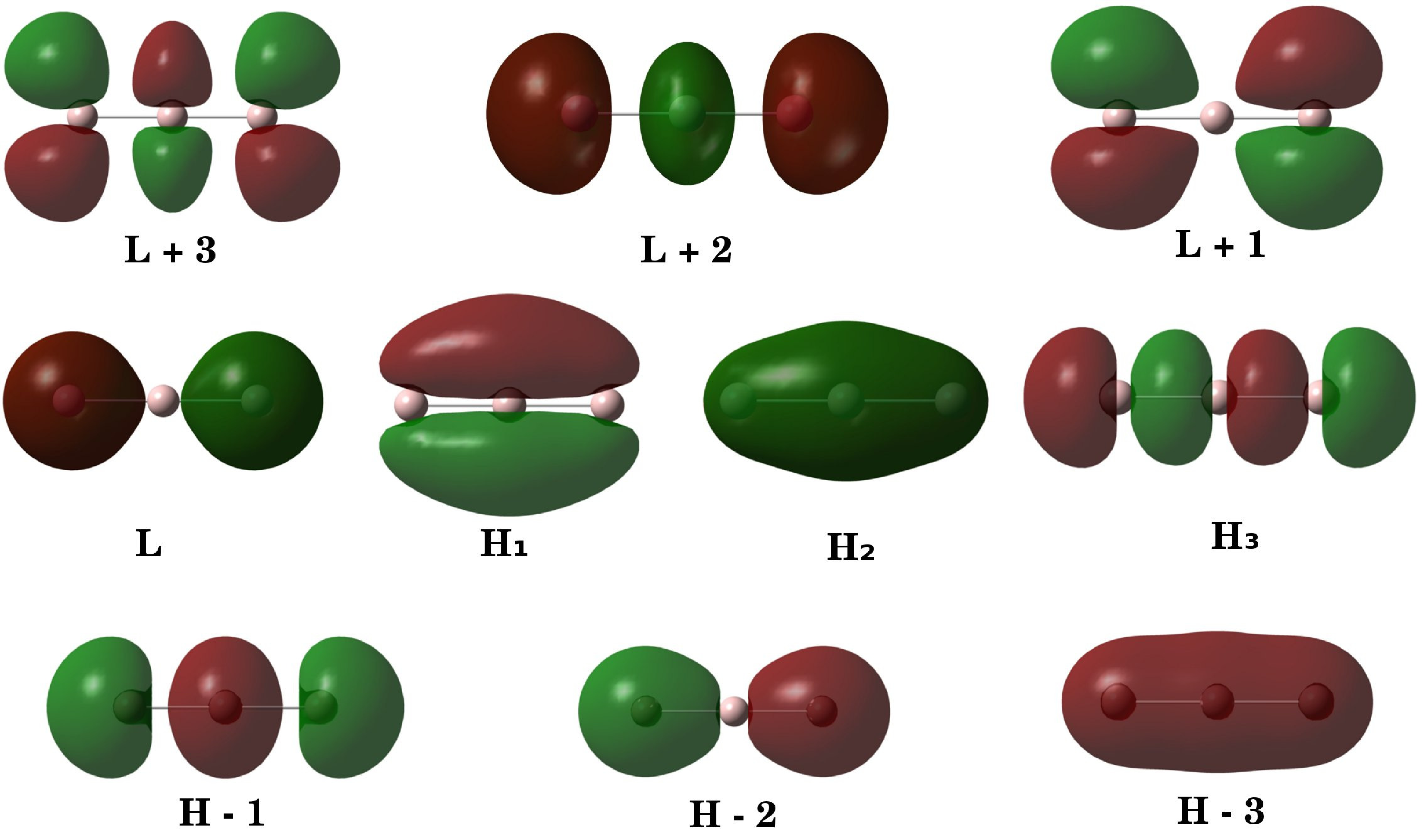}
 \caption{Molecular orbitals of linear aluminum trimer. $H$
 and $L$ stands for HOMO and LUMO respectively, and $H_{1}$, $H_{2}$,
 and $H_{3}$ are singly occupied molecular orbitals.}
 \end{figure}

 \begin{figure}
 \includegraphics[width=7cm]{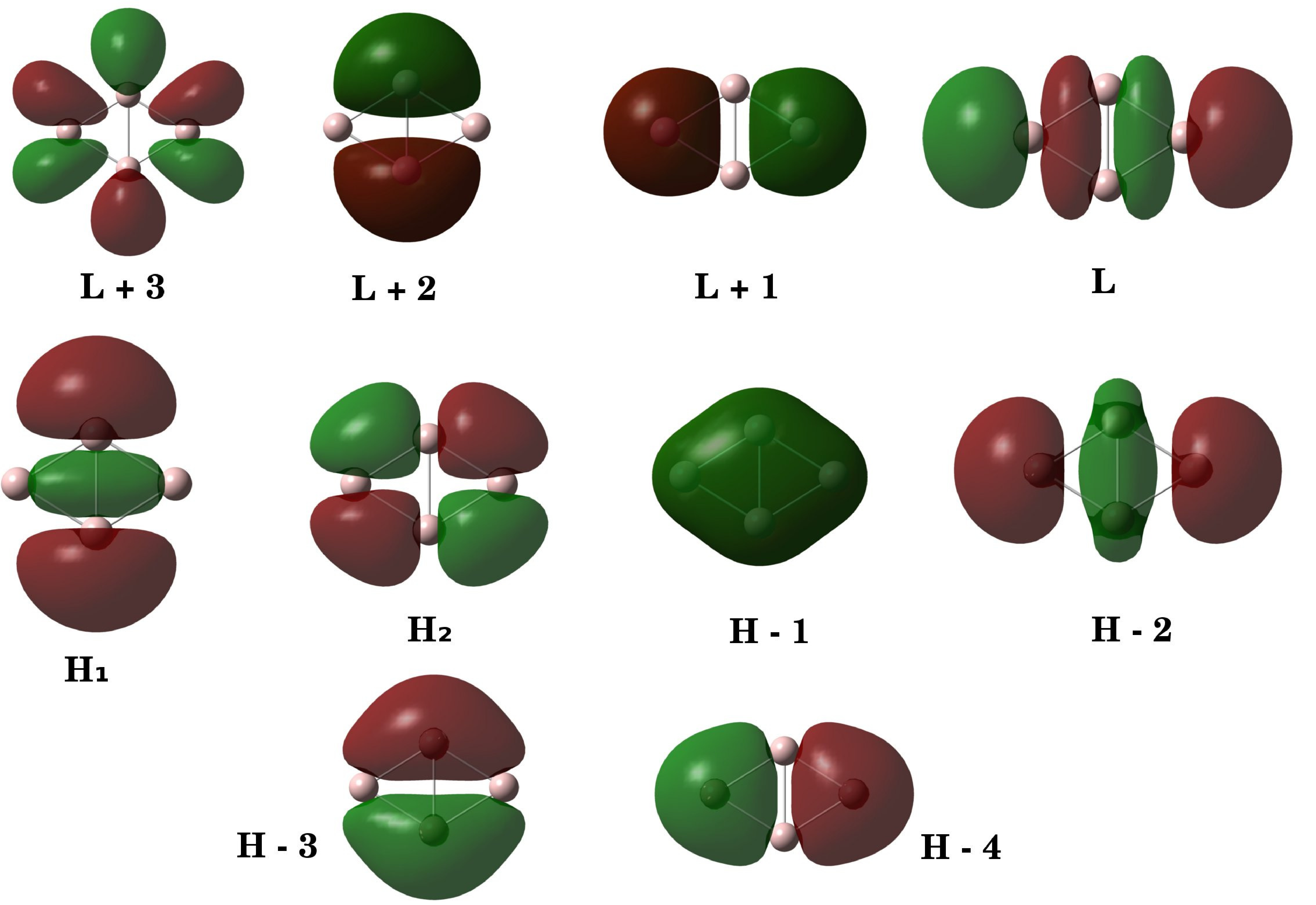}
 \caption{Molecular orbitals of rhombus-shaped aluminum tetramer.
 $H$ and $L$ stands for HOMO and LUMO respectively, and $H_{1}$
 and $H_{2}$ are singly occupied molecular orbitals.}
 \end{figure}

 \begin{figure}
 \includegraphics[width=7cm]{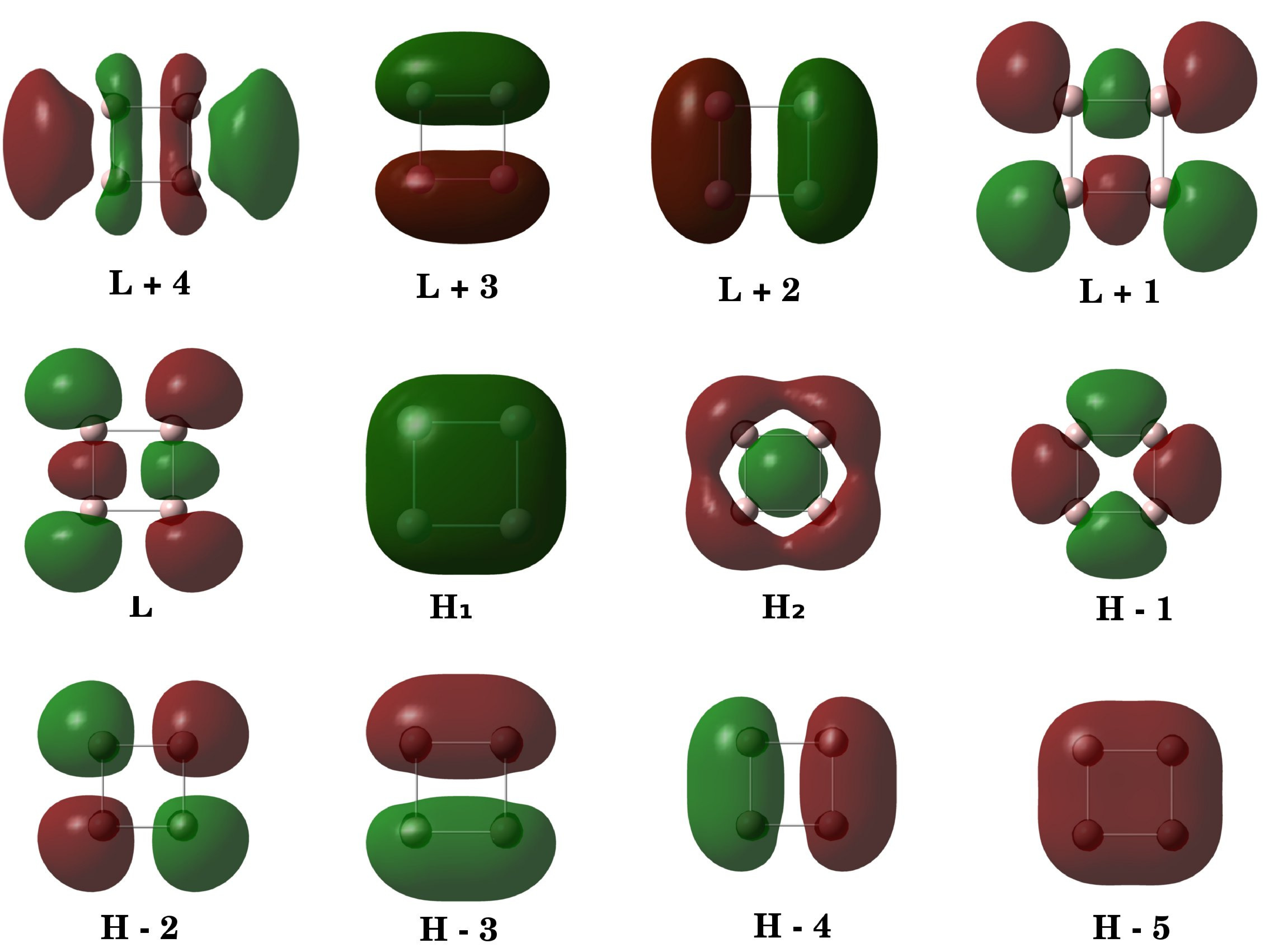}
 \caption{Molecular orbitals of square-shaped aluminum tetramer.
 $H$ and $L$ stands for HOMO and LUMO respectively, and $H_{1}$
 and $H_{2}$ are singly occupied molecular orbitals.}
 \end{figure}

 \begin{figure}
 \includegraphics[width=7cm]{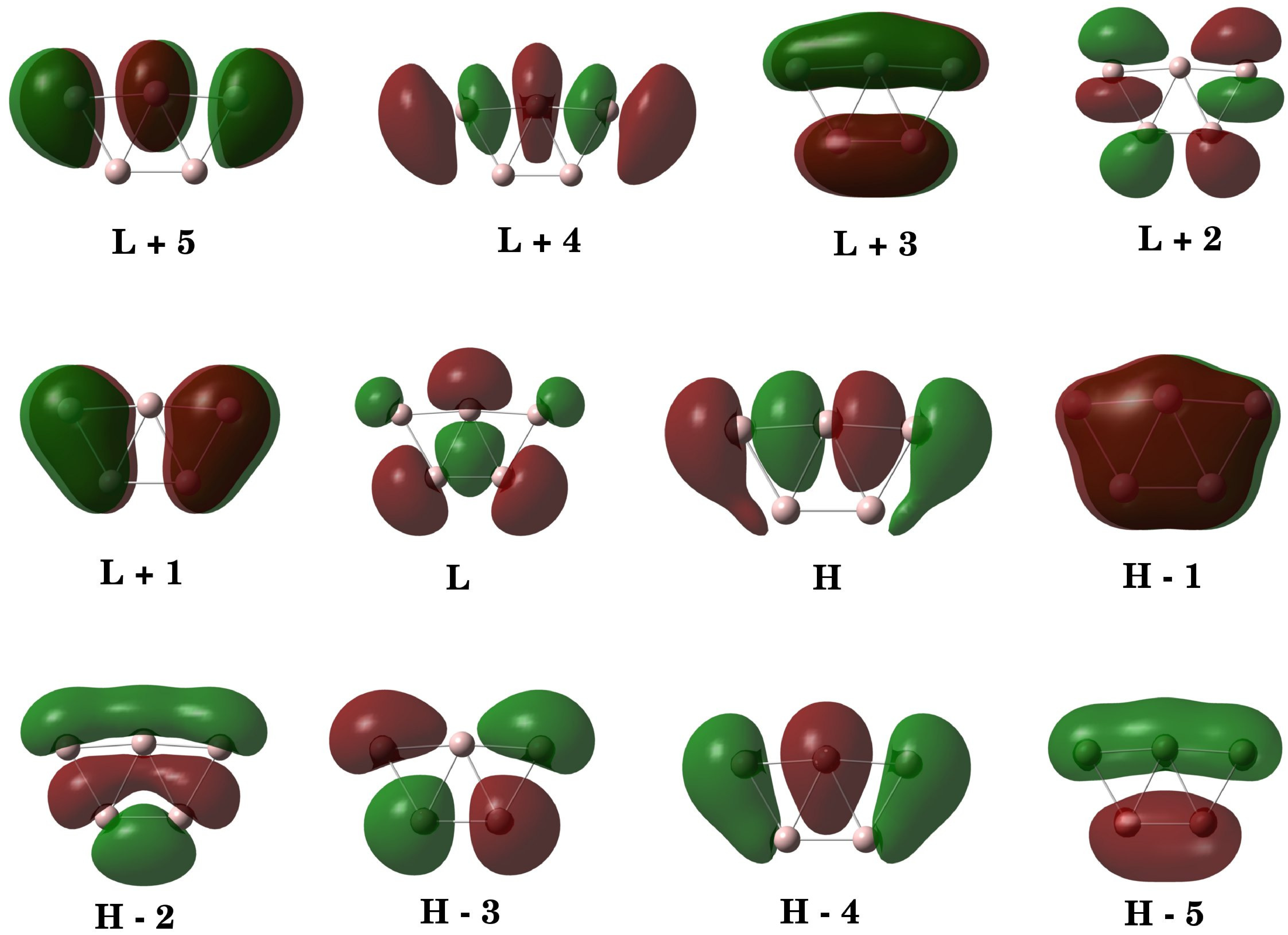}
 \caption{Molecular orbitals of pentagonal aluminum pentamer.
 $H$ and $L$ stands for HOMO and LUMO respectively.}
 \end{figure}

 \begin{figure}
 \includegraphics[width=7cm]{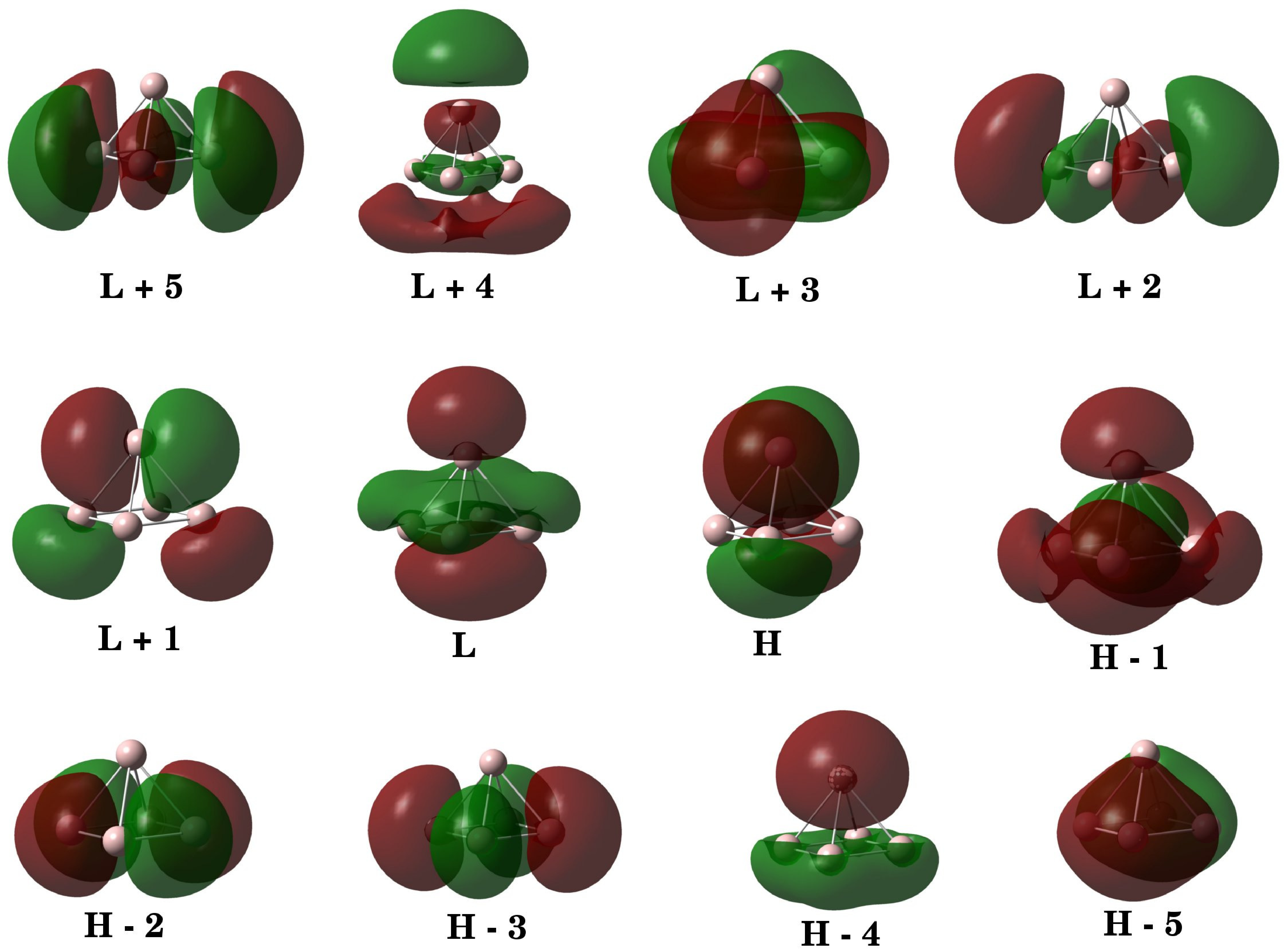}
 \caption{Molecular orbitals of pyramidal aluminum pentamer.
 $H$ and $L$ stands for HOMO and LUMO respectively.}
 \end{figure}

 \FloatBarrier
% \LTcapwidth=1.0\textwidth
\begin{table*}
\caption{ Excitation energies ($E$) and many-particle wave functions of excited
states corresponding to the peaks in the linear absorption spectrum
of Al$_{2}$% (\emph{cf}. Fig. \ref{fig:plot-al2-linear})
, along with the oscillator strength ($f_{12}$) of the transitions. Longitudinal
and transverse polarization corresponds to the absorption due to light
polarized along and perpendicular to the molecular axis respectively.
In the wave function, the bracketed numbers are the CI coefficients
of a given electronic configuration. Symbols $H_{1}$,$H_{2}$ denote
SOMOs discussed earlier, and $H$, and $L$, denote HOMO and LUMO
orbitals respectively. $HF$ denotes the Hartree-Fock configuration. }  

\begin{tabular}{cccccl}
Peak &  $E$ (eV) &	Symmetry &$f_{12}$ & Polarization & Wave Function  \tabularnewline \hline 
GS\footnotemark[1]
 &  & $^{3}B_{2u}$  &  &  & $|H_{1}^{1},\, H_{2}^{1}\rangle$ (0.9096)\tabularnewline
 &  &   &  &  & $|H-1\rightarrow H_{1};H_{2}\rightarrow L\rangle$(0.1139)\tabularnewline
 &  &   &  &  & $|H-2\rightarrow L;H-1\rightarrow L+2\rangle$(0.0889)\tabularnewline
 &  &   &  &  & \tabularnewline
I  & 1.96 & $^{3}B_{3g}$ & 0.1027 & longitudinal & $|H_{2}\rightarrow L+1\rangle$(0.8120) \tabularnewline
 &  &  &  & & $|H-1\rightarrow H_{1}\rangle$(0.3685) \tabularnewline
 &  &  &  & & \tabularnewline
II  & 3.17 & $^{3}B_{3g}$ &0.1249 & longitudinal & $|H-1\rightarrow H_{1}\rangle$(0.6172) \tabularnewline
 &  &  &  & & $|H_{1}\rightarrow L+3\rangle$(0.4068) \tabularnewline
 &  &  &  & & $|H_{1}\rightarrow L;H-1\rightarrow L\rangle$(0.3190) \tabularnewline
 &  &  &  & & \tabularnewline
III  & 4.47 &$^{3}A_{g}$ & 0.5149 & transverse & $|H_{2}\rightarrow L+4\rangle$(0.8313) \tabularnewline
 &  &  &  & & $|H_{2}\rightarrow L+6\rangle$(0.2024) \tabularnewline
 &  &  &  & & \tabularnewline
IV  & 4.99 &$^{3}B_{3g}$ & 5.4531 & longitudinal & $|H_{1}\rightarrow L+3\rangle$(0.7353) \tabularnewline
 &  &  &  & & $|H-1\rightarrow H_{1}\rangle$(0.4104) \tabularnewline
 &  &  &  & & \tabularnewline
V & 6.31 & $^{3}A_{g}$ & 0.2554 & transverse & $|H_{2}\rightarrow L+6\rangle$(0.4683) \tabularnewline
 &  &  &  & & $|H-1\rightarrow L+1\rangle$(0.3894) \tabularnewline
 &  &  &  & & $|H-1\rightarrow L;H_{2}\rightarrow L+2\rangle$(0.3886) \tabularnewline
 &  &  &  & \tabularnewline
VI & 7.17 & $^{3}A_{g}$ & 0.1549 & transverse & $|H_{2}\rightarrow L+2;H-1\rightarrow L\rangle$(0.4782) \tabularnewline
 &  &  &  & & $|H-1\rightarrow L+1\rangle$(0.4327) \tabularnewline
 &  &  &  & & $|H_{1}\rightarrow L;H_{2}\rightarrow L+8\rangle$(0.3867) \tabularnewline
 &  &  &  & & \tabularnewline
VII & 7.79 &  $^{3}A_{g}$ & 1.2530 & transverse & $|H-1\rightarrow H_{1};H_{2}\rightarrow L+3\rangle$(0.4833) \tabularnewline
 &  &  &  & & $|H_{1}\rightarrow L+7\rangle$(0.3917) \tabularnewline
 &  &  &  & & $|H_{1}\rightarrow L;H_{2}\rightarrow L+8\rangle$(0.3791) \tabularnewline
 &  &  &  & & \tabularnewline
VIII & 8.05 &  $^{3}B_{1g}$ & 3.5391 & transverse & $|H-2\rightarrow L\rangle$(0.5316) \tabularnewline
 &  &  &  & & $|H-1\rightarrow L+2\rangle$(0.3756) \tabularnewline
 &  &  &  & & $|H_{1}\rightarrow L+8\rangle$(0.3531) \tabularnewline
 & 8.10 & $^{3}A_{g}$ & 1.1418 & transverse & $|H-1\rightarrow H_{1};H_{2}\rightarrow L+3\rangle$(0.4788) \tabularnewline
 &  &  &  & & $|H_{2}\rightarrow L+6\rangle$(0.4095) \tabularnewline
 &  &  &  & & \tabularnewline
IX & 8.87 & $^{3}B_{1g}$ & 0.7044 & transverse & $|H_{1}\rightarrow L+11\rangle$(0.5061) \tabularnewline
 &  &  &  & & $|H_{1}\rightarrow L;H_{2}\rightarrow L+7\rangle$(0.4162) \tabularnewline
 & 8.95 & $^{3}A_{g}$ & 0.6872 & transverse & $|H_{1}\rightarrow L+7\rangle$(0.4932) \tabularnewline
 &  &  &  & & $|H_{2}\rightarrow L;H_{1}\rightarrow L+8\rangle$(0.4414) \tabularnewline
 &  &  &  & & $|H_{1}\rightarrow L+4;H-1\rightarrow L+1\rangle$(0.3262) \tabularnewline
\label{Tab:table_al2_lin}
\end{tabular}
\footnotetext[1]{$GS$ does not correspond to any peak, rather it corresponds to the
ground state wavefunction of Al$_{2}$ isomer.}%}

\end{table*}

% \LTcapwidth=1.0\textwidth
\begin{table*}
\caption{Excitation energies ($E$) and many-particle wave functions of excited
states corresponding to the peaks in the linear absorption spectrum
of metastable Al$_{2}$% (\emph{cf}. Fig. \ref{fig:plot-al2-linear})
, along with the oscillator strength ($f_{12}$) of the transitions. Longitudinal
and transverse polarization corresponds to the absorption due to light
polarized along and perpendicular to the molecular axis respectively.
In the wave function, the bracketed numbers are the CI coefficients
of a given electronic configuration. Symbols $H_{1}$,$H_{2}$ denote
SOMOs discussed earlier, and $H$, and $L$, denote HOMO and LUMO
orbitals respectively. $HF$ denotes the Hartree-Fock configuration. }  

\begin{tabular}{cccccl}
Peak &  $E$ (eV) &	Symmetry &$f_{12}$ & Polarization & Wave Function  \tabularnewline \hline 
GS\footnotemark[2]
 &  & $^{3}B_{3g}$  &  &  & $|H_{1}^{1},\, H_{2}^{1}\rangle$ (0.8975)\tabularnewline
 &  &   &  &  & $|H-1\rightarrow L;H-1\rightarrow L  \rangle$(0.1418)\tabularnewline
 &  &   &  &  & $|H-1\rightarrow L;H_{1}\rightarrow L + 1\rangle$(0.1146)\tabularnewline
 &  &   &  &  & \tabularnewline
I  & 2.29 & $^{3}A_{u}$ & 0.0283 & longitudinal & $|H - 1 \rightarrow L\rangle$(0.6598) \tabularnewline
 &  &  &  & & $|H_{1}\rightarrow L + 1 \rangle$(0.4276) \tabularnewline
 &  &  &  & & $|H_{2}\rightarrow L + 1 \rangle$(0.4276) \tabularnewline
 &  &  &  & & \tabularnewline
II  & 3.26 & $^{3}A_{u}$ &0.0350 & longitudinal & $|H-1\rightarrow L \rangle$(0.7659) \tabularnewline
 &  &  &  & & $|H_{1}\rightarrow L \rangle$(0.3137) \tabularnewline
 &  &  &  & & \tabularnewline
III  & 4.40 &$^{3}B_{2u,3u}$ & 0.0469 & transverse & $|H - 1 \rightarrow L+1\rangle$(0.5540) \tabularnewline
 &  &  &  & & $|H_{2}\rightarrow L+1;H - 1 \rightarrow H_{1} \rangle$(0.4827) \tabularnewline
 &  &  &  & & \tabularnewline
IV  & 4.67  &$^{3}B_{2u,3u}$& 0.1769 & transverse & $|H_{1}\rightarrow L+3\rangle$(0.5073) \tabularnewline
 &  &  &  & & $|H-1\rightarrow L + 1 \rangle$(0.5030) \tabularnewline
 &  &  &  & & \tabularnewline
V & 5.17 & $^{3}A_{u}$ & 5.8490 & longitudinal & $|H - 1 \rightarrow L \rangle$(0.7286) \tabularnewline
 &  &  &  & & $|H_{1}\rightarrow L+1\rangle$(0.3078) \tabularnewline
 &  &  &  & \tabularnewline
VI & 5.75 & $^{3}B_{2u,3u}$ & 0.1549 & transverse & $|H-1\rightarrow H_{2} \rangle$(0.5354) \tabularnewline
 &  &  &  & & $|H-1\rightarrow L+1\rangle$(0.4847) \tabularnewline
 &  &  &  & & \tabularnewline
VII & 6.24 &  $^{3}B_{2u,3u}$ & 0.2361 & transverse & $|H-1\rightarrow L+1\rangle$(0.5856) \tabularnewline
 &  &  &  & & $|H - 1 \rightarrow L;H - 1 \rightarrow H_{1} \rangle$(0.3432) \tabularnewline
 &  &  &  & & \tabularnewline
VIII & 6.79 &  $^{3}B_{2u,3u}$ & 0.0659 & transverse & $|H_{1}\rightarrow L;H_{2} \rightarrow L + 5 \rangle$(0.4766) \tabularnewline
 &  &  &  & & $|H_{1} \rightarrow L; H - 1 \rightarrow L + 2 \rangle$(0.4333) \tabularnewline
 &  &  &  & & \tabularnewline
IX & 7.73 & $^{3}A_{u}$ & 0.5428 & longitudinal & $|H - 1 \rightarrow L+3 \rangle$(0.6484) \tabularnewline
 &  &  &  & & $|H_{1}\rightarrow L+1;H - 1 \rightarrow L \rangle$(0.2333) \tabularnewline
 &  &  &  & & \tabularnewline
X & 8.13 & $^{3}B_{2u,3u}$ & 3.6959 & transverse & $|H - 1 \rightarrow L+1 \rangle$(0.4767) \tabularnewline
 &  &  &  & & $|H_{2}\rightarrow L+4 \rangle$(0.4052) \tabularnewline
 &  &  &  & & \tabularnewline
XI & 8.49 & $^{3}B_{2u,3u}$ & 1.1382 & transverse & $|H_{1} \rightarrow L; H-1 \rightarrow L + 2 \rangle$(0.4727) \tabularnewline
 &  &  &  & & $|H - 2 \rightarrow H_{1} \rangle$(0.3364) \tabularnewline
\label{Tab:table_al2_meta_lin}
\end{tabular}
\footnotetext[2]{$GS$ does not correspond to any peak, rather it corresponds to the
ground state wavefunction of metastable Al$_{2}$ isomer.}%}

\end{table*}

\begin{table*}
 \caption{Excitation energies ($E$) and many-particle wave functions of excited
states corresponding to the peaks in the linear absorption spectrum
of Al$_{3}$ equilateral triangle isomer% (\emph{cf}. Fig. \ref{fig:plot-al3-equil})
, along with the oscillator strength ($f_{12}$) of the transitions.
In-plane and transverse polarization corresponds to the absorption
due to light polarized in and perpendicular to the plane of the triangular
isomer respectively. In the wave function, the bracketed numbers are
the CI coefficients of a given electronic configuration. Symbols $H$
and $L$, denote HOMO (singly occupied, in this case) and LUMO orbitals
respectively. $HF$ denotes the Hartree-Fock configuration. } \label{Tab:table_al3_equil_tri}

\begin{tabular}{cccccl}
Peak &  $E$ (eV) & Symmetry &  $f_{12}$ & Polarization & Wave Function  \tabularnewline \hline 
GS\footnotemark[3]
 &  &  $^{2}A_{1}$  & &  & $|HF\rangle$ (0.8373)\tabularnewline
 &  &  & & & $|H-2\rightarrow L+5\rangle$(0.1329)\tabularnewline
 &  &  & &  & \tabularnewline
I  & 3.42 & $^{2}B_{2}$  &0.0376 & in-plane & $|H-3\rightarrow L+5\rangle$(0.2908) \tabularnewline
 &  &  &  & & $|H-2\rightarrow L+1\rangle$(0.2439) \tabularnewline
 & 3.54 & $^{2}A_{1}$  & 0.1080 & in-plane & $|H-2\rightarrow L+5\rangle$(0.3686) \tabularnewline
 &  &  &  & &  $|H-2\rightarrow H\rangle$(0.3403) \tabularnewline
 &  &  &  & & \tabularnewline
II  & 5.61 & $^{2}A_{1}$  & 0.2565 & in-plane & $|H-2\rightarrow L+5;H-1\rightarrow L+5\rangle$(0.4854) \tabularnewline
 &  &  &  & & $|H\rightarrow L+1;H-1\rightarrow L+1\rangle$(0.4476) \tabularnewline
 &  &  &  & & \tabularnewline
III  & 5.87 & $^{2}B_{1}$  & 0.3413 & transverse & $|H-3\rightarrow L+2\rangle$(0.2915) \tabularnewline
 &  &  &  & & $|H-2\rightarrow L\rangle$(0.2842) \tabularnewline
 &  &  &  & & \tabularnewline
IV & 6.53 & $^{2}A_{1}$  & 6.3289 & in-plane & $|H\rightarrow L+6\rangle$(0.4044) \tabularnewline
 &  &  &  & & $|H-3\rightarrow L+1\rangle$(0.3965) \tabularnewline
 &  &  &  & & $|H-2\rightarrow L+5\rangle$(0.3158) \tabularnewline
 & 6.53 & $^{2}B_{2}$  & 5.7925 & in-plane & $|H\rightarrow L+4\rangle$(0.3842) \tabularnewline
 &  &  &  & & $|H-3\rightarrow L+5\rangle$(0.2834) \tabularnewline
 &  &  &  & & $|H-4\rightarrow L+1\rangle$(0.2256) \tabularnewline
 &  &  &  & \tabularnewline
V & 6.96 & $^{2}B_{1}$  & 0.4145 & transverse & $|H-2\rightarrow L\rangle$(0.3140) \tabularnewline
 &  &  &  & & $|H-3\rightarrow L+2\rangle$(0.2626) \tabularnewline
 &  &  &  & & \tabularnewline
VI & 7.50 & $^{2}B_{2}$  & 0.9430 & in-plane & $|H-2\rightarrow L+1;H\rightarrow L+5\rangle$(0.3136) \tabularnewline
 &  &  &  & & $|H-3\rightarrow L+5\rangle$(0.2864) \tabularnewline
 & 7.57 & $^{2}A_{1}$  & 0.8630 & in-plane & $|H\rightarrow L+5;H-3\rightarrow L+1\rangle$(0.3838) \tabularnewline
 &  &  &  & & $|H-3\rightarrow L+1\rangle$(0.2651) \tabularnewline
 &  &  &  & & $|H-2\rightarrow L+5\rangle$(0.2590) \tabularnewline
\end{tabular}
\footnotetext[3]{$GS$ does not correspond to any peak, rather it corresponds to the
ground state wavefunction of Al$_{3}$ equilateral triangle isomer.}

\end{table*}%  

\begin{table*}
 \caption{Excitation energies ($E$) and many-particle wave functions of excited
states corresponding to the peaks in the linear absorption spectrum
of Al$_{3}$ isosceles triangle isomer% (\emph{cf}. Fig. \ref{fig:plot-al3-iso}),
, along with the oscillator strength ($f_{12}$) of the transitions.
In-plane and transverse polarization corresponds to the absorption
due to light polarized in and perpendicular to the plane of the triangular
isomer respectively. In the wave function, the bracketed numbers are
the CI coefficients of a given electronic configuration. Symbols $H_{1}$,
$H_{2}$ and $H_{3}$ denote SOMOs discussed earlier, $H$ and $L$,
denote HOMO and LUMO orbitals respectively. } \label{Tab:table_al3_iso_tri}

\begin{tabular}{cccccl}
Peak &  $E$ (eV) & Symmetry & $f_{12}$ & Polarization & Wave Function  \tabularnewline \hline 
GS\footnotemark[4]
 &  &   $^{4}A_{1}$  & &   & $|H_{1}^{1},H_{2}^{1},H_{3}^{1}\rangle$ (0.8670)\tabularnewline
 &  &  &  & & $|H-1\rightarrow L+10\rangle$(0.1213)\tabularnewline
 &  &  &  & & \tabularnewline
I & 2.37 & $^{4}A_{2}$  & 0.0358 & in-plane & $|H_{1}\rightarrow L+1;H_{3}\rightarrow L+2\rangle$(0.7066) \tabularnewline
 &  &  &  & & $|H-1\rightarrow L+1;H_{1}\rightarrow L\rangle$(0.4052) \tabularnewline
 &  &  &  & & \tabularnewline
II & 3.06 & $^{4}B_{1}$  &  0.0992 & in-plane & $|H_{3}\rightarrow H_{2};H-2\rightarrow L\rangle$(0.4691) \tabularnewline
 &  &  &  & & $|H-1\rightarrow L+1;H_{3}\rightarrow H_{2}\rangle$(0.4070) \tabularnewline
 &  &  &  & & \tabularnewline
III  & 3.45 & $^{4}A_{2}$  & 0.0967 & in-plane & $|H_{1}\rightarrow L+3\rangle$(0.5566) \tabularnewline
 &  &  &  & & $|H-1\rightarrow L+1;H_{1}\rightarrow L\rangle$(0.5209) \tabularnewline
 &  &  &  & & \tabularnewline
IV  & 4.11 & $^{4}B_{1}$  & 0.3208 & in-plane & $|H_{1}\rightarrow L+4\rangle$(0.6038) \tabularnewline
 &  &  &  & & $|H_{3}\rightarrow L+1;H-2\rightarrow L\rangle$(0.5272) \tabularnewline
 &  &  &  & & \tabularnewline
V & 4.83 & $^{4}A_{2}$  &  0.2242 & in-plane & $|H_{1}\rightarrow L+1;H-2\rightarrow L+1\rangle$(0.5321) \tabularnewline
 &  &  &  & & $|H_{1}\rightarrow L+5\rangle$(0.2611) \tabularnewline
 &  &  &  & & \tabularnewline
VI & 5.76 & $^{4}A_{2}$  &  5.0792 & in-plane & $|H-1\rightarrow L+1;H_{3}\rightarrow L\rangle$(0.3479) \tabularnewline
 &  &  &  & & $|H-3\rightarrow L+1;H_{1}\rightarrow L\rangle$(0.2875) \tabularnewline
 &  &  &  & & $|H_{2}\rightarrow L+1;H_{1}\rightarrow L+3\rangle$(0.2800) \tabularnewline
 & 5.85 & $^{4}B_{1}$  & 0.8553 & in-plane & $|H_{3}\rightarrow L+1;H-2\rightarrow L\rangle$(0.4081) \tabularnewline
 &  &  &  & & $|H-1\rightarrow L;H_{3}\rightarrow L\rangle$(0.2400) \tabularnewline
 &  &  &  & & \tabularnewline
VII & 5.95 & $^{4}A_{2}$  &  1.7094 & in-plane & $|H-1\rightarrow L+2\rangle$(0.3296) \tabularnewline
 &  &  &  & & $|H-1\rightarrow L+1;H_{3}\rightarrow L\rangle$(0.3138) \tabularnewline
 & 6.15 & $^{4}B_{1}$  & 0.7827 & in-plane & $|H_{1}\rightarrow L+7\rangle$(0.7827) \tabularnewline
 &  &  & & & \tabularnewline
VIII & 6.68 & $^{4}B_{1}$  & 1.7774 & in-plane & $|H_{1}\rightarrow L+10\rangle$(0.4548) \tabularnewline
 &  &  & & & $|H_{2}\rightarrow L+1;H_{1}\rightarrow L+6\rangle$(0.2705) \tabularnewline
 &  &  & & & $|H_{1}\rightarrow L+6\rangle$(0.2447) \tabularnewline
\end{tabular}
\footnotetext[4]{$GS$ does not correspond to any peak, rather it corresponds to the
ground state wavefunction of Al$_{3}$ isosceles triangle isomer.}

\end{table*}% 

\begin{table*}
 \caption{Excitation energies ($E$) and many-particle wave functions of excited
states corresponding to the peaks in the linear absorption spectrum
of Al$_{3}$ linear isomer, % (\emph{cf}. Fig. \ref{fig:plot-al3-lin}),
along with the oscillator strength ($f_{12}$) of the transitions.
Longitudinal and transverse polarization corresponds to the absorption
due to light polarized along and perpendicular to the axis of the
linear isomer respectively. In the wave function, the bracketed numbers
are the CI coefficients of a given electronic configuration. Symbols
$H_{1}$, $H_{2}$ and $H_{3}$ denote SOMOs discussed earlier, $H$
and $L$, denote HOMO and LUMO orbitals respectively. $HF$ denotes
the Hartree-Fock configuration. } \label{Tab:table_al3_lin_tri}

\begin{tabular}{cccccl}
Peak &  $E$ (eV) & Symmetry & $f_{12}$ & Polarization & Wave Function  \tabularnewline \hline 
GS\footnotemark[5]
 &  & $^{4}A_{u}$  &  &  & $|H_{1}^{1},H_{2}^{1},H_{3}^{1}\rangle$ (0.8010)\tabularnewline
 &  &  & &  & $|H-3\rightarrow H_{1};H_{3}\rightarrow L\rangle$(0.1913)\tabularnewline
 &  &  & &  & \tabularnewline
I & 1.24 & $^{4}B_{3g}$  &  0.0317 & longitudinal & $|H_{2}\rightarrow L+1\rangle$(0.6602) \tabularnewline
 &  &  & &  & $|H-1\rightarrow H_{3}\rangle$(0.3636) \tabularnewline
 &  &  & & & \tabularnewline
II & 2.25 & $^{4}B_{3g}$  &0.0489 & longitudinal & $|H-1\rightarrow H_{3}\rangle$(0.6856) \tabularnewline
 &  &  & &  & $|H-2\rightarrow H_{1}\rangle$(0.3230) \tabularnewline
 &  &  & &  & \tabularnewline
III  & 4.01 & $^{4}B_{3g}$  &0.9019 & longitudinal & $|H-2\rightarrow H_{1}\rangle$(0.5249) \tabularnewline
 &  &  & & & $|H-1\rightarrow H_{3}\rangle$(0.3471) \tabularnewline
 &  &  & & & \tabularnewline
IV  & 4.43 & $^{4}B_{3g}$  & 2.8593 & longitudinal & $|H-1\rightarrow H_{3}\rangle$(0.4070) \tabularnewline
 &  &  & & & $|H-1\rightarrow L+4;H_{2}\rightarrow L+6\rangle$(0.2409) \tabularnewline
 & 4.47 & $^{4}B_{1g,2g}$  & 0.0960 & transverse & $|H_{2}\rightarrow L+2\rangle$(0.5402) \tabularnewline
 &  &  & & & $|H-1\rightarrow H_{3};H_{2}\rightarrow L+6\rangle$(0.3068) \tabularnewline
 &  &  & & & \tabularnewline
V & 4.62 & $^{4}B_{3g}$  & 5.1747 & longitudinal & $|H-1\rightarrow H_{3}\rangle$(0.4600) \tabularnewline
 &  &  & & & $|H-1\rightarrow L+4;H_{2}\rightarrow L+6\rangle$(0.2862) \tabularnewline
 &  &  &  & & \tabularnewline
VI & 5.29 & $^{4}B_{1g,2g}$  & 0.1070 & transverse & $|H_{2}\rightarrow L+5\rangle$(0.4951) \tabularnewline
 &  &  & &  & $|H-1\rightarrow H_{3};H-1\rightarrow L+1\rangle$(0.3284) \tabularnewline
 &  &  & & & $|H-1\rightarrow L+3\rangle$(0.3091) \tabularnewline
 &  &  &  & & \tabularnewline
VII & 5.83 & $^{4}B_{3g}$  & 0.1412 & longitudinal & $|H-1\rightarrow L+2;H_{1}\rightarrow L\rangle$(0.6637) \tabularnewline
 &  &  &  & &  $|H-2\rightarrow H_{1}\rangle$(0.2225) \tabularnewline
 &  &  &  & & $|H-1\rightarrow H_{3}\rangle$(0.2073) \tabularnewline
 &  &  &  & & \tabularnewline
VIII & 6.31 & $^{4}B_{3g}$  & 0.0459 & longitudinal & $|H_{1}\rightarrow L+6;H_{3}\rightarrow L\rangle$(0.5099) \tabularnewline
 &  &  &  & & $|H_{1}\rightarrow L;H_{3}\rightarrow L+6\rangle$(0.2706) \tabularnewline
 & 6.37 & $^{4}B_{1g,2g}$  & 0.0740 & transverse & $|H-1\rightarrow L+3\rangle$(0.3989) \tabularnewline
 &  &  & &  & $|H-1\rightarrow H_{2};H_{3}\rightarrow L+6\rangle$(0.2266) \tabularnewline
 &  &  & & & \tabularnewline
IX & 6.89 & $^{4}B_{3g}$  &0.1311 & longitudinal & $|H-5\rightarrow L+6\rangle$(0.3920) \tabularnewline
 &  &  & &  & $|H_{1}\rightarrow L+4;H_{3}\rightarrow L+6\rangle$(0.3086) \tabularnewline
\end{tabular}
\footnotetext[5]{$GS$ does not correspond to any peak, rather it corresponds to the
ground state wavefunction of Al$_{3}$ linear triangle isomer.}

\end{table*} % 

\begin{table*}
 \caption{Excitation energies ($E$) and many-particle wave functions of excited
states corresponding to the peaks in the linear absorption spectrum
of Al$_{4}$ rhombus isomer, % (\emph{cf}. Fig. \ref{fig:plot-al4-rho}),
along with the oscillator strength ($f_{12}$) of the transitions.
In-plane and transverse polarization corresponds to the absorption
due to light polarized in and perpendicular to the plane of the rhombus
isomer respectively. In the wave function, the bracketed numbers are
the CI coefficients of a given electronic configuration. Symbols $H_{1}$,$H_{2}$
denote SOMOs discussed earlier, and $H$, and $L$, denote HOMO and
LUMO orbitals respectively.} \label{Tab:table_al4_rho}

\begin{tabular}{cccccl}

 Peak &  $E$ (eV) & Symmetry & $f_{12}$ & Polarization & Wave Function  \tabularnewline \hline 
GS\footnotemark[6]
 &  & $^{3}B_{2g}$  & &  & $|H_{1}^{1},H_{2}^{1}\rangle$ (0.8724)\tabularnewline
 &  & &  &  & $|H-3\rightarrow L;H-3\rightarrow L\rangle$(0.1050)\tabularnewline
 &  & &  &  & \tabularnewline
I  & 1.07 & $^{3}B_{1u}$  & 0.0247 & transverse & $|H_{1}\rightarrow L+1\rangle$(0.8489) \tabularnewline
 &  &  &  & & $|H-2\rightarrow L+5\rangle$(0.1601) \tabularnewline
 &  &  &  & & \tabularnewline
II  & 2.31 & $^{3}B_{3u}$  & 0.3087 & in-plane & $|H-2\rightarrow H_{1}\rangle$(0.7645) \tabularnewline
 &  &  &  & & $|H_{2}\rightarrow L+1\rangle$(0.3113) \tabularnewline
 &  &  &  & & \tabularnewline
III  & 4.67 & $^{3}B_{3u}$  & 0.5709 & in-plane & $|H-2\rightarrow L;H-1\rightarrow L+3\rangle$(0.6036) \tabularnewline
 &  &  &  & & $|H-1\rightarrow L+3\rangle$(0.4213) \tabularnewline
 &  &  &  & & $|H_{1}\rightarrow L+7\rangle$(0.3113) \tabularnewline
 &  &  &  & & \tabularnewline
IV  & 4.88 & $^{3}A_{u}$  & 0.9622 & in-plane & $|H-1\rightarrow L;H-1\rightarrow L+3\rangle$(0.6036) \tabularnewline
 &  &  &  & & $|H-3\rightarrow L\rangle$(0.4699) \tabularnewline
 &  &  &  & & \tabularnewline
V & 5.51 & $^{3}B_{3u}$  & 3.8316 & in-plane & $|H-3\rightarrow L+4\rangle$(0.7378) \tabularnewline
 &  &  &  & & $|H-2\rightarrow H_{1}\rangle$(0.2161) \tabularnewline
 &  &  &  & & \tabularnewline
VI & 5.84 & $^{3}A_{u}$  & 0.4900 & in-plane & $|H-2\rightarrow L+3\rangle$(0.3889) \tabularnewline
 &  &  &  & & $|H-2\rightarrow L;H-3\rightarrow L\rangle$(0.3758) \tabularnewline
 &  &  &  & & $|H-3\rightarrow L\rangle$(0.3594) \tabularnewline
 &  &  &  & & $|H-1\rightarrow L;H-1\rightarrow L+3\rangle$(0.3591) \tabularnewline
 &  &  &  & & \tabularnewline
VII & 6.01 & $^{3}B_{1u}$  & 0.5332 & transverse & $|H_{2}\rightarrow L+7\rangle$(0.7268) \tabularnewline
 &  &  &  & & $|H-3\rightarrow L+2\rangle$(0.3050) \tabularnewline
 &  &  &  & & \tabularnewline
VIII & 6.20 & $^{3}A_{u}$  & 0.7477 & in-plane & $|H-2\rightarrow L+3\rangle$(0.5195) \tabularnewline
 &  &  &  & & $|H-2\rightarrow L;H-3\rightarrow L\rangle$(0.4189) \tabularnewline
 &  &  &  & & \tabularnewline
IX & 6.51 & $^{3}B_{1u}$  & 0.2928 & transverse & $|H-3\rightarrow L+2\rangle$(0.7001) \tabularnewline
 &  &  &  & & $|H-2\rightarrow H_{1};H-1\rightarrow L+2\rangle$(0.2232) \tabularnewline
 &  &  &  & & $|H-2\rightarrow L;H-3\rightarrow L+2\rangle$(0.2070) \tabularnewline
 &  &  &  & & \tabularnewline
X & 6.92 & $^{3}B_{1u}$  & 0.6053 & transverse & $|H-3\rightarrow L+2\rangle$(0.5144) \tabularnewline
 &  &  &  & & $|H-2\rightarrow L;H-3\rightarrow L+2\rangle$(0.3549) \tabularnewline
 &  &  &  & & $|H-2\rightarrow L+5\rangle$(0.2676) \tabularnewline
 &  &  &  & & \tabularnewline
XI & 7.31 & $^{3}B_{1u}$  & 0.4328 & transverse & $|H-2\rightarrow L+5\rangle$(0.4033) \tabularnewline
 &  &  &  & & $|H-3\rightarrow L;H-1\rightarrow L+1\rangle$(0.3787) \tabularnewline
 &  &  &  & & \tabularnewline
XII & 7.76 & $^{3}B_{3u}$  & 2.7450 & in-plane & $|H_{1}\rightarrow L+8\rangle$(0.4387) \tabularnewline
 &  &  &  & & $|H_{1}\rightarrow L+1;H-1\rightarrow L+2\rangle$(0.3435) \tabularnewline
\end{tabular}
\footnotetext[6]{$GS$ does not correspond to any peak, rather it corresponds to the
ground state wavefunction of Al$_{4}$ rhombus isomer.}

\end{table*}% 

\begin{table*}
 \caption{Excitation energies ($E$) and many-particle wave functions of excited
states corresponding to the peaks in the linear absorption spectrum
of Al$_{4}$ square isomer, % (\emph{cf}. Fig. \ref{fig:plot-al4-sqr}),
along with the oscillator strength ($f_{12}$) of the transitions.
In-plane and transverse polarization corresponds to the absorption
due to light polarized in and perpendicular to the plane of the rhombus
isomer respectively. In the wave function, the bracketed numbers are
the CI coefficients of a given electronic configuration. Symbols $H_{1}$,$H_{2}$
denote SOMOs discussed earlier, and $H$, and $L$, denote HOMO and
LUMO orbitals respectively.} \label{Tab:table_al4_sqr}

\begin{tabular}{cccccl}
Peak &  $E$ (eV) & Symmetry & $f_{12}$ & Polarization & Wave Function  \tabularnewline \hline 
GS\footnotemark[7]
 &   & $^{3}B_{3u}$  &  &  & $|H_{1}^{1},H_{2}^{1}\rangle$(0.8525)\tabularnewline
 &  &  &  & & $|H_{1}\rightarrow L;H-2\rightarrow L\rangle$(0.0972)\tabularnewline
 &  &  &  & & \tabularnewline
I  & 2.08 &  $^{3}B_{1g,2g}$  & 0.0278 & in-plane & $|H-1\rightarrow L\rangle$(0.7191) \tabularnewline
 &  &  & &  & $|H-1\rightarrow H_{1};H_{2}\rightarrow L+1\rangle$(0.2645) \tabularnewline
 &  &  & &  & $|H_{2}\rightarrow L+1\rangle$(0.2536) \tabularnewline
 &  &  & &  & $|H_{1}\rightarrow L\rangle$(0.2443) \tabularnewline
 &  &  &  & &  \tabularnewline
II  & 2.68 & $^{3}B_{1g,2g}$  & 0.0301 & in-plane & $|H_{2}\rightarrow L+1\rangle$(0.4757) \tabularnewline
 &  &  &  & & $|H-1\rightarrow L\rangle$(0.4358) \tabularnewline
 &  &  &  & & $|H-1\rightarrow H_{1};H_{2}\rightarrow L+1\rangle$(0.3608)\tabularnewline
 &  &  &  & & \tabularnewline
III  & 4.19 & $^{3}B_{1g,2g}$  & 0.3420 & in-plane & $|H-2\rightarrow L\rangle$(0.5889) \tabularnewline
 &  &  &  & & $|H-1\rightarrow L+2\rangle$(0.4283) \tabularnewline
 &  &  &  & & $|H_{1}\rightarrow L\rangle$(0.2329) \tabularnewline
 &  &  &  & & \tabularnewline
IV  & 4.92 & $^{3}B_{1g,2g}$  & 0.1131 & in-plane & $|H_{1}\rightarrow L+2\rangle$(0.5780) \tabularnewline
 &  &  &  & &  $|H-1\rightarrow L+2\rangle$(0.4083) \tabularnewline
 &  &  &  & & $|H-2\rightarrow L\rangle$(0.3198) \tabularnewline
 &  &  &  & & \tabularnewline
V & 5.17 & $^{3}A_{g}$  & 0.1238 & transverse & $|H-2\rightarrow L;H_{1}\rightarrow L+1\rangle$(0.3693)\tabularnewline
 &  &  &  & & $|H-2\rightarrow L;H_{1}\rightarrow L+1\rangle$(0.3692)\tabularnewline
 & 5.33 & $^{3}B_{1g,2g}$  &  0.2470 & in-plane & $|H-2\rightarrow H_{1};H-2\rightarrow L\rangle$(0.5193)\tabularnewline
 &  &  &  & & $|H-1\rightarrow L+2\rangle$(0.3915)\tabularnewline
 &  &  &  & & $|H-2\rightarrow L+2\rangle$(0.3335)\tabularnewline
 &  &  &  & & \tabularnewline
VI & 5.85 & $^{3}B_{1g,2g}$  & 1.2446 & in-plane & $|H-2\rightarrow L+2\rangle$(0.7184) \tabularnewline
 &  &  &  & & $|H-1\rightarrow H_{1};H-2\rightarrow L+2\rangle$(0.2587) \tabularnewline
 &  &  &  & & $|H-1\rightarrow L+2\rangle$(0.2579) \tabularnewline
 &  &  &  & & \tabularnewline
VII & 6.55 & $^{3}B_{1g,2g}$  &  3.7894 & in-plane & $|H-2\rightarrow L+2\rangle$(0.5706) \tabularnewline
 &  &  &  & & $|H-1\rightarrow H_{1};H-2\rightarrow L+2\rangle$(0.4089) \tabularnewline
 &  &  &  & & $|H-1\rightarrow L+2\rangle$(0.3325) \tabularnewline
 & 6.58 & $^{3}A_{g}$  & 0.2634 & transverse & $|H_{1}\rightarrow L+1;H-2\rightarrow L\rangle$(0.4375) \tabularnewline
 &  &  &  & & $|H_{1}\rightarrow L+1;H-2\rightarrow L\rangle$(0.4375) \tabularnewline
 &  &  &  & & $|H-2\rightarrow L+3\rangle$(0.4183) \tabularnewline
 &  &  &  & & \tabularnewline
VIII & 6.87 & $^{3}B_{1g,2g}$  & 2.9702 & in-plane & $|H-2\rightarrow L+2\rangle$(0.5100) \tabularnewline
 &  &  &  & & $|H-1\rightarrow L+2\rangle$(0.3495) \tabularnewline
 & 6.93 & $^{3}A_{g}$  & 0.2483 & transverse & $|H_{1}\rightarrow L+1;H-2\rightarrow L\rangle$(0.3558) \tabularnewline
 &  &  &  & &  $|H_{1}\rightarrow L+1;H-2\rightarrow L\rangle$(0.3558) \tabularnewline
 &  &  &  & & $|H-2\rightarrow L+3\rangle$(0.2929) \tabularnewline
 &  &  &  & & \tabularnewline
IX & 7.22 & $^{3}B_{1g,2g}$  & 1.4267 & in-plane & $|H-1\rightarrow H_{1};H-2\rightarrow L+2\rangle$(0.4039) \tabularnewline
 &  &  &  & & $|H-3\rightarrow H_{1}\rangle$(0.2900) \tabularnewline
 &  &  &  & & $|H-1\rightarrow H_{1};H-1\rightarrow L+2\rangle$(0.2848)\tabularnewline
\end{tabular}
\footnotetext[7]{$GS$ does not correspond to any peak, rather it corresponds to the
ground state wavefunction of Al$_{4}$ square isomer.}

\end{table*}% 

\begin{table*}
 \caption{Excitation energies ($E$) and many-particle wave functions of excited
states corresponding to the peaks in the linear absorption spectrum
of Al$_{5}$ pentagonal isomer, % (\emph{cf}. Fig. \ref{fig:plot-al5-penta}),
along with the oscillator strength ($f_{12}$) of the transitions.
In-plane and transverse polarization corresponds to the absorption
due to light polarized in and perpendicular to the plane of the pentagonal
isomer respectively. In the wave function, the bracketed numbers are
the CI coefficients of a given electronic configuration. Symbols $H$
and $L$, denote HOMO and LUMO orbitals respectively.} \label{Tab:table_al5_pentagon}

\begin{tabular}{cccccl}
Peak &  $E$ (eV) & Symmetry & $f_{12}$ & Polarization & Wave Function  \tabularnewline \hline 
GS\footnotemark[8]
 &  &    $^{2}A_{1}$ &  & & $|(H-2)^{1}\rangle$ (0.8679)\tabularnewline
 &  &  &  & & $|H-2\rightarrow L+1;H\rightarrow L+2\rangle$(0.1045)\tabularnewline
 &  &  &  & & \tabularnewline
I  & 1.03 & $^{2}B_{2}$  & 0.0195 & in-plane & $|H-1\rightarrow L\rangle$(0.8635) \tabularnewline
 &  &  &  & &  $|H-1\rightarrow L;H\rightarrow L+3\rangle$(0.0880) \tabularnewline
 &  &  &  & & \tabularnewline
II  & 2.38 & $^{2}B_{2}$  & 0.0219 & in-plane & $|H-3\rightarrow H-2\rangle$(0.8560) \tabularnewline
 &  &  &  & & $|H-1\rightarrow L+4\rangle$(0.1387) \tabularnewline
 &  &  &  & & \tabularnewline
III  & 3.90 & $^{2}B_{1}$  & 0.1042 & transverse & $|H\rightarrow L+4\rangle$(0.8387) \tabularnewline
 &  &  &  & & $|H\rightarrow L;H-1\rightarrow L+2\rangle$(0.1944) \tabularnewline
 &  & $^{2}A_{1}$  & 0.3362 & in-plane & $|H-4\rightarrow L\rangle$(0.8140) \tabularnewline
 &  &  &  & & $|H-2\rightarrow L+9\rangle$(0.1841) \tabularnewline
 &  &  &  & & \tabularnewline
IV  & 4.16 & $^{2}B_{2}$  & 1.3144 & in-plane & $|H-1\rightarrow L+4\rangle$(0.7276) \tabularnewline
 &  &  &  & & $|H-1\rightarrow L+5\rangle$(0.4478) \tabularnewline
 &  &  &  & & \tabularnewline
V & 4.42 & $^{2}B_{2}$  & 3.3339 & in-plane & $|H-1\rightarrow L+5\rangle$(0.7096) \tabularnewline
 &  &  &  & & $|H-1\rightarrow L+4\rangle$(0.4490) \tabularnewline
 &  &  &  & & $|H-1\rightarrow L+9\rangle$(0.1535) \tabularnewline
 &  &  &  & & \tabularnewline
VI & 4.78 & $^{2}A_{1}$  & 1.0471 & in-plane & $|H-2\rightarrow L+9\rangle$(0.7992) \tabularnewline
 &  &  &  & & $|H-2\rightarrow L;H\rightarrow L+6\rangle$(0.2058) \tabularnewline
 &  &  &  & & \tabularnewline
VII & 5.46 & $^{2}B_{1}$  & 1.1014 & transverse & $|H\rightarrow L+13\rangle$(0.8156) \tabularnewline
 &  &  &  & & $|H\rightarrow L;H-2\rightarrow L\rangle$(0.1708) \tabularnewline
 &  &  &  & & \tabularnewline
VIII & 6.37 & $^{2}B_{2}$  & 0.1270 & in-plane & $|H-3\rightarrow L\rangle$(0.7632) \tabularnewline
 &  &  &  & & \tabularnewline
IX & 6.73 & $^{2}B_{2}$  & 0.7104 & in-plane & $|H-3\rightarrow L\rangle$(0.7370) \tabularnewline
 &  &  &  & & $|H\rightarrow L+1\rangle$(0.3698) \tabularnewline
 &  &  &  & & $|H-1\rightarrow L;H\rightarrow L+3\rangle$(0.1225) \tabularnewline
 &  &  &  & & \tabularnewline
X & 7.49 & $^{2}A_{1}$  & 0.3989 & in-plane & $|H\rightarrow L+3\rangle$(0.5087) \tabularnewline
 &  &  &  & & $|H-2\rightarrow L+16\rangle$(0.3508) \tabularnewline
 &  &  &  & & $|H\rightarrow L;H-1\rightarrow L+1\rangle$(0.2937) \tabularnewline
\end{tabular}
\footnotetext[8]{$GS$ does not correspond to any peak, rather it corresponds to the
ground state wavefunction of Al$_{5}$ pentagonal isomer.}

\end{table*} % 

\begin{table*}
 \caption{Excitation energies ($E$) and many-particle wave functions of excited
states corresponding to the peaks in the linear absorption spectrum
of Al$_{5}$ pyramid isomer, % (\emph{cf}. Fig. \ref{fig:plot-al5-pyra}),
along with the oscillator strength ($f_{12}$) of the transitions.
In the wave function, the bracketed numbers are the CI coefficients
of a given electronic configuration. Symbols $H$ and $L$, denote
HOMO and LUMO orbitals respectively.} \label{Tab:table_al5_pyramid}

\begin{tabular}{cccccl}
 Peak &  $E$ (eV) & Symmetry & $f_{12}$ & Polarization & Wave Function  \tabularnewline \hline 
GS\footnotemark[9]%
 &  & $^{2}A_{1}$  &  &  & $|(H-2)^{1}\rangle$ (0.8591)\tabularnewline
 &  &  &  & & $|H-3\rightarrow L+1;H-3\rightarrow L+1\rangle$(0.1138)\tabularnewline
 &  &  &  & & \tabularnewline
I  & 1.72 & $^{2}B_{2}$  & 0.0046 & y & $|H-3\rightarrow L+1\rangle$(0.6849) \tabularnewline
 &  &  &  & & $|H-2\rightarrow L+1\rangle$(0.2887) \tabularnewline
 & 1.75 & $^{2}A_{1}$  & 0.0521 & z & $|H\rightarrow L+3\rangle$(0.2887) \tabularnewline
 &  &  &  & & \tabularnewline
II  & 2.21 & $^{2}B_{2}$  & 0.0296 & y & $|H-3\rightarrow L+1\rangle$(0.7170) \tabularnewline
 &  &  &  & & $|H-2\rightarrow L+2\rangle$(0.3402) \tabularnewline
 &  &  &  & & $|H-3\rightarrow L+2\rangle$(0.2290) \tabularnewline
 &  &  &  & & \tabularnewline
III  & 2.55 & $^{2}A_{1}$  & 0.0477 & z & $|H\rightarrow L+3\rangle$(0.5390) \tabularnewline
 &  &  &  & & $|H-4\rightarrow H-2\rangle$(0.1296) \tabularnewline
 &  &  &  & & \tabularnewline
IV & 3.46 & $^{2}B_{2}$  & 0.0399 & y & $|H-3\rightarrow L;H-2\rightarrow L+1\rangle$(0.6131) \tabularnewline
 &  &  &  & & $|H-3\rightarrow L+2\rangle$(0.4975) \tabularnewline
  & 3.48 & $^{2}A_{1}$  & 0.0769 & z & $|H-4\rightarrow H-2\rangle$(0.7340) \tabularnewline
 &  &  &  & & $|H-4\rightarrow L\rangle$(0.3735) \tabularnewline
 &  &  &  & & \tabularnewline
V & 4.04 & $^{2}B_{1}$  & 0.6432 & x & $|H\rightarrow L+7\rangle$(0.5929) \tabularnewline
 &  &  &  & & $|H\rightarrow L+4\rangle$(0.4432) \tabularnewline
 & 4.22 & $^{2}B_{2}$  & 3.0735 & y & $|H-3\rightarrow L+2\rangle$(0.8272) \tabularnewline
 &  &  &  & & $|H-3\rightarrow L+1\rangle$(0.1580) \tabularnewline
 &  &  &  & & \tabularnewline
VI & 4.74 & $^{2}B_{1}$  & 0.3474 & x & $|H\rightarrow L\rangle$(0.7617) \tabularnewline
 &  &  &  & & $|H\rightarrow L+7\rangle$(0.2542) \tabularnewline
 &  &  &  & & \tabularnewline
VII & 5.08 & $^{2}A_{1}$  & 0.5494 & z & $|H-2\rightarrow L;H\rightarrow L+2\rangle$(0.5540) \tabularnewline
 &  &  &  & & $|H-4\rightarrow L\rangle$(0.4833) \tabularnewline
 &  &  &  & & \tabularnewline
VIII & 5.26 & $^{2}A_{1}$  & 0.3175 & z & $|H-2\rightarrow L;H\rightarrow L+5\rangle$(0.6251) \tabularnewline
 &  &  &  & & $|H-4\rightarrow L\rangle$(0.3902) \tabularnewline
 & 5.27 & $^{2}B_{1}$  & 0.1267 & x & $|H-6\rightarrow H-2;H\rightarrow L+1\rangle$(0.6056) \tabularnewline
 &  &  &  & & $|H\rightarrow L\rangle$(0.3242) \tabularnewline
 &  &  &  & & \tabularnewline
IX & 5.56 & $^{2}B_{1}$  & 0.1384 & x & $|H\rightarrow L+11\rangle$(0.7819) \tabularnewline
 &  &  &  & & $|H\rightarrow L+13\rangle$(0.3051) \tabularnewline
 &  &  &  & & \tabularnewline
X & 6.00 & $^{2}B_{1}$  & 1.0052 & x & $|H\rightarrow L+13\rangle$(0.8132) \tabularnewline
 &  &  &  & &  $|H\rightarrow L+11\rangle$(0.1852) \tabularnewline
\end{tabular}
\footnotetext[9]{$GS$ does not correspond to any peak, rather it corresponds to the
ground state wavefunction of Al$_{5}$ pyramid isomer. }

\end{table*}
%  \bibliographystyle{unsrt}
% \bibliography{smallal} 
\end{document}